\documentclass[aoas]{imsart}

\RequirePackage{amsthm,amsmath,amsfonts,amssymb}
\RequirePackage[authoryear]{natbib}
\RequirePackage[colorlinks,citecolor=blue,urlcolor=blue]{hyperref}
\RequirePackage{graphicx}
\graphicspath{{figure/}}

\usepackage{enumerate}

\usepackage{bm}
\usepackage{dsfont}
\usepackage{amssymb}
\usepackage{amstext}
\usepackage{graphics}
\usepackage{graphicx}
\usepackage{amsthm}
\usepackage{amsmath}
\usepackage{mathtools}
\usepackage{enumitem}
\usepackage{color}

\usepackage{booktabs}
\usepackage{longtable}
\usepackage{array}
\usepackage{multirow}
\usepackage{wrapfig}
\usepackage{float}
\usepackage{colortbl}
\usepackage{pdflscape}
\usepackage{tabu}
\usepackage{threeparttable}
\usepackage{threeparttablex}
\usepackage[normalem]{ulem}
\usepackage{makecell}
\usepackage{xcolor}

\usepackage{tikz}
\usepackage{extarrows}
\usetikzlibrary{fadings}
\usetikzlibrary{patterns}
\usetikzlibrary{shadows}
\usetikzlibrary{shadows.blur}
\usetikzlibrary{shapes}

\usepackage{subcaption}

\startlocaldefs

\definecolor{bleudefrance}{rgb}{0.19, 0.55, 0.91}
\newcommand{\suppref}[1]{\textcolor{bleudefrance}{#1}}

\makeatletter
\def\namedlabel#1#2{\begingroup
    #2%
    \def\@currentlabel{#2}%
    \phantomsection\label{#1}\endgroup
}
\makeatother

\newcommand*{\fullref}[1]{\hyperref[{#1}]{\autoref*{#1} of \nameref*{#1}}}

\theoremstyle{plain}

\newtheorem{ps}{Proposition}[section]

\definecolor{amber}{rgb}{1.0, 0.75, 0.0}
\definecolor{ao}{rgb}{0.0, 0.0, 1.0}

\AtBeginDocument{}
\makeatletter
\newsavebox{\mybox}\newsavebox{\mysim}
\newcommand{\distas}[1]{%
    \savebox{\mybox}{\hbox{\kern3pt$\scriptstyle#1$\kern3pt}}%
    \savebox{\mysim}{\hbox{$\sim$}}%
    \mathbin{\overset{\text{#1}}{\kern\z@\resizebox{\wd\mybox}{\ht\mysim}{$\sim$}}}%
}
\makeatother
\newcommand{\defas}{\coloneqq}

\newcommand*{\cbrk}[1]{\left\{ #1 \right\}} 

\DeclareMathOperator*{\argmin}{arg\,min}

\renewcommand*{\cal}[1]{\mathcal{#1}}

\newcommand{\calS}{\mathcal{S}}

\newcommand{\calG}{\mathcal{G}}

\newcommand{\dsR}{\mathds{R}}

\newcommand{\vphi}{\varphi}

\newcommand*{\tl}[1]{\tilde{#1}}
\newcommand*{\h}[1]{\hat{#1}}
\newcommand*{\ck}[1]{\check{#1}}

\newcommand{\var}{\operatorname{var}}

\newcommand{\cov}{\operatorname{cov}}

\DeclarePairedDelimiterX{\infdivx}[2]{(}{)}{%
    #1\;\delimsize\|\;#2%
}

\renewcommand*{\vec}[1]{\bm{#1}}
\newcommand*{\mat}[1]{\mathbf{#1}}

\newcommand*{\inv}{^{-1}}

\newcommand*{\tpose}{^{T}}

\newcommand{\prob}{\mathds{P}}
\newcommand{\Exp}{\mathds{E}}

\newcommand{\waspace}[1][2]{\cbx W_#1}

\newcommand{\wassdsq}[2]{d_W^2\left(#1, #2\right)}
\newcommand{\wasslog}[1][{\lambda}]{\log_{#1}}
\newcommand{\wassexp}[1][{\lambda}]{\exp_{#1}}

\newcommand{\fave}{\oplus}

\newcommand{\ginv}{^-} 
\newcommand{\qf}[1][]{F_{#1}\ginv} 
\newcommand{\cdf}[1][]{{F_{#1}}} 
\newcommand{\emp}[1]{\ck{#1}}
\newcommand{\empqf}[1][]{\emp F_{#1}\ginv} 
\newcommand{\empcdf}[1][]{{\emp F_{#1}}} 

\newcommand{\smplaveinline}[2][n]{{#1}^{-1} \sum_{#2=1}^#1} 

\newcommand{\pushforward}{_\#}
\newcommand{\id}{\mathop{\text{id}}}

\newcommand{\norm}[1]{\left\lVert#1\right\rVert}
\DeclarePairedDelimiterX{\abs}[1]{\lvert}{\rvert}{#1}

\newcommand{\innerL}[3][\lambda]{\left\langle #2,\  #3 \right\rangle_{#1}}
\newcommand{\normL}[2][\lambda]{\left\lVert#2\right\rVert_{#1}}

\newcommand{\logit}{\operatorname{logit}}
\newcommand{\ttv}[1]{\texttt{#1}} 

\newcommand{\upup}[1]{{#1}^\uparrow}

\newcommand{\functional}{\cbx{f}}

\newcommand{\frechet}{{Fr\'echet }}
\newcommand{\matern}{{Mat\'ern }}

\DeclareMathAlphabet{\cbx}{U}{BOONDOX-calo}{m}{n}
\SetMathAlphabet{\cbx}{bold}{U}{BOONDOX-calo}{b}{n}
\DeclareMathAlphabet{\cbxb}{U}{BOONDOX-calo}{b}{n}

\endlocaldefs

\begin{document}

\begin{frontmatter}
\title{Spatial Prediction of Local Soil Erosion Distribution \\ in the Wasserstein Space}

\begin{aug}

\author[A]{\fnms{Jiaming}~\snm{Qiu}\ead[label=e1]{jqiu3@fredhutch.org}\orcid{0000-0002-2174-0337}},
\author[B]{\fnms{Xiongtao}~\snm{Dai}\ead[label=e2]{xdai@berkeley.edu}\orcid{0000-0002-6996-5930}},
\author[C]{\fnms{Zhengyuan}~\snm{Zhu}\ead[label=e3]{zhuz@iastate.edu}\orcid{0000-0002-2266-0646}}
\and
\author[D]{\fnms{Shuiqing}~\snm{Yin}\ead[label=e4]{yinshuiqing@bnu.edu.cn}\orcid{0000-0001-6914-6006}}
\address[A]{Public Health Sciences Division, Fred Hutchinson Cancer Center \printead[presep={,\ }]{e1}}

\address[B]{Division of Biostatistics, the University of California, Berkeley \printead[presep={,\ }]{e2}}

\address[C]{Department of Statistics, Iowa State University \printead[presep={,\ }]{e3}}

\address[D]{State Key Laboratory of Earth Surface Processes and Hazards Risk Governance (ESPHR),\\ Faculty of Geographical Science, Beijing Normal University \printead[presep={,\ }]{e4}}

\end{aug}

\begin{abstract}

Obtaining precise erosion measurements requires costly fieldwork, making it infeasible to directly survey large domains such as a province or river basin. To extend fieldwork results across such extensive domains, we propose a novel spatial prediction method that treats local erosion distributions as objects in the Wasserstein space. These distributions are mapped into square-integrable trajectories and represented via basis expansion, forming a multivariate random field that captures spatial dependence. By applying local regression and Kriging in this representation, our approach flexibly models and predicts erosion distributions at arbitrary locations. This framework improves prediction for functionals of the distribution, such as the mean and exceedance probabilities. Simulation studies demonstrate that the proposed method outperforms a misspecified parametric alternative and existing \frechet regression approaches. We illustrate the approach with a detailed erosion analysis in Shaanxi province, China, where local measurements from surveyed watersheds are extended to predict erosion distributions across the entire province using covariates such as land use and elevation.
\end{abstract}

\begin{keyword}
\kwd{Wasserstein space}
\kwd{Functional data analysis}
\kwd{Kriging}
\kwd{Zero-inflated model}
\end{keyword}

\end{frontmatter}


\section{Introduction}

\subsection{Erosion in Shaanxi Province of China}\label{sec:data}

Soil erosion has been extensively studied to prevent associated risks such as soil degradation and flooding.
The extent of soil erosion could drastically vary even within a small local watershed, making laborious fieldwork a necessity for a precise assessment.
To wisely allocate treatment resources, it is vital to characterize the erosion profile of sampling units, e.g., small local watersheds, and highlight those with a substantial portion of highly eroded lands.
Our data consists of detailed field survey data collected from limited watersheds. The goal is to predict the local erosion distributions in the more extensive domain of the entire province.

\begin{figure}
    \centering
    \includegraphics[width=0.85\textwidth]{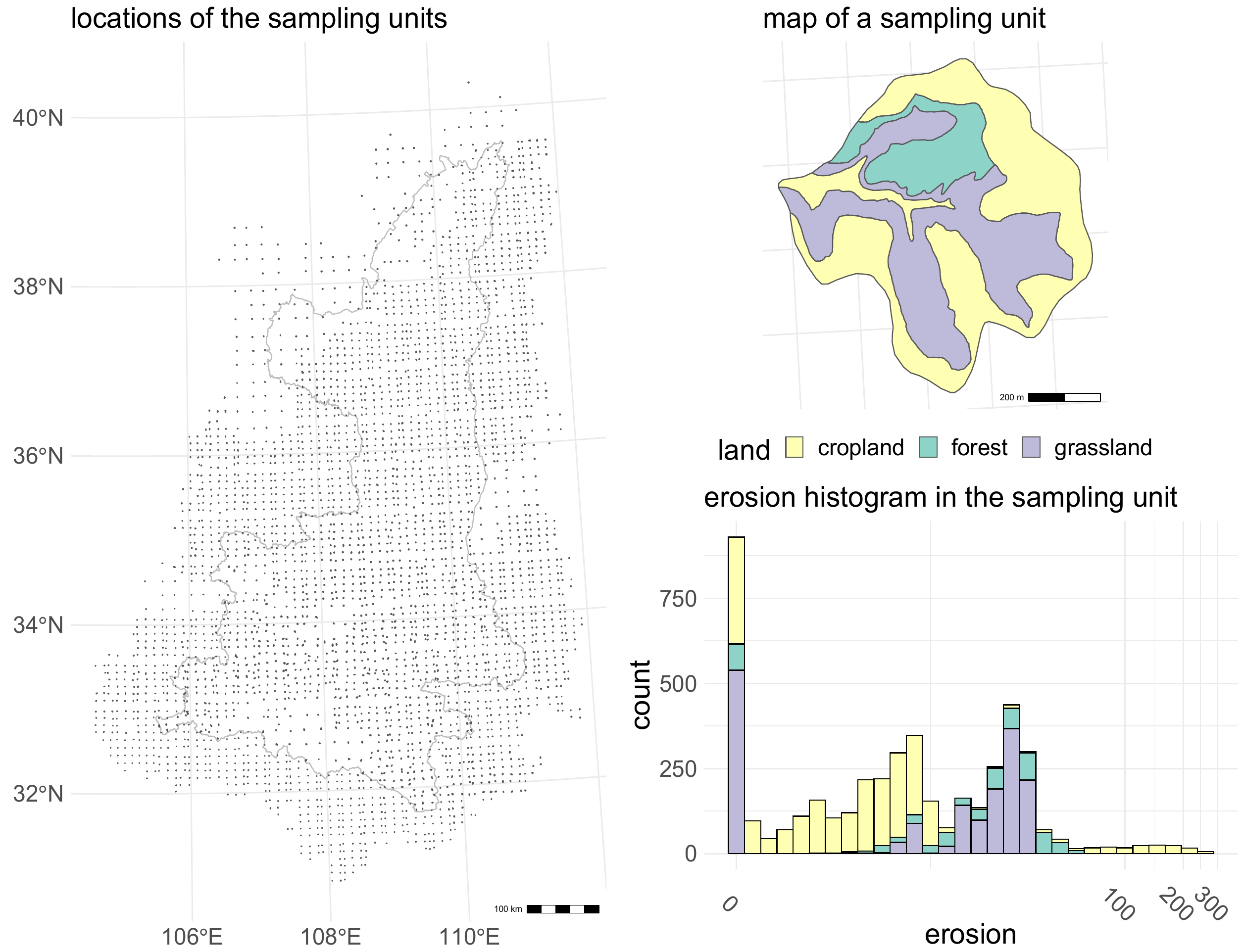}
    \caption{
        Left panel: the province boundary (polygon) and surveyed sampling units (dots). 
        Right panels: details in an example sampling unit with a surface area of 0.45 $\text{km}^2$, which recorded $4506$ individual soil loss values at 10-meter resolution obtained in fieldwork. 
        The upper right panel maps its land type. The lower right panel shows the distributions of local erosion values. 
        Note that there are a non-negligible portion of zeros, and that the erosion distributions substantially differ among the different land types.
    }
    \label{fig:sample.unit}
\end{figure}

The Shaanxi province in China is one of the most eroded provinces along the Yellow River basin. As part of a nationwide erosion study \citep{zhen:13}, 3628 \emph{sampling units} were surveyed across the convex hull of the Shaanxi province with a 30 km buffer (the left panel of \autoref{fig:sample.unit}).
Each sampling unit represents a small watershed sized approximately one square kilometer (e.g., the upper right panel of \autoref{fig:sample.unit}).
All sampling units were surveyed by combing topographical information and fieldwork to provide a local erosion map with a resolution of $10 \times 10$ meters, where for each pixel a soil loss value was calculated according to the empirical Soil Loss Equation \citep[see, e.g., ][]{liu:02}
\begin{align}\label{def:soil_loss}
    & \quad \ttv{A} =
    \ttv{R} \times \ttv{K} \times \ttv{L} \times \ttv{S} \times 
    \ttv{B} \times \ttv{E} \times \ttv{T}. 
\end{align}
The resulting soil erosion \ttv{A} is the product of local contributing factors, including the rainfall erodability $\ttv{R}$; the soil erodability $\ttv{K}$; the length $\ttv{L}$ and steepness $\ttv{S}$ of the slope; and the biological, engineering, and tillage practice factors ($\ttv{B}$, $\ttv{E}$, and $\ttv{T}$); the measurement unit of \ttv{A} is tons per hectare per year (t/ha/yr).
Pooling together soil losses of all pixels within the sampling unit provides a probability distribution representing the local erosion characteristic (e.g., the lower right panel of \autoref{fig:sample.unit}).
See \cite{zhen:13,liu:13,yin:18-1} for details of this survey.

One of our goals is to use the local erosion distribution to identify the severely eroded watersheds so that treatment resources can be intelligently allocated.
However, precise measurements of many factors in \eqref{def:soil_loss} (e.g.~\ttv{L}, \ttv{S}, \ttv{B}, \ttv{E}, \ttv{T}) are unavailable without fieldwork and high-resolution topography, making it infeasible to directly survey all watersheds of interest. The existing data only cover 1\% of the surface area in the province, leaving the massive unsurveyed region relying on interpolation/prediction.

\subsection{Spatially Distributed Distributions}\label{sec:introMethod}

With the local watershed as the primary unit, our analysis targets its probability distribution of erosion.
This perspective encompasses the classical approach \citep[e.g.,][]{yin:18-1} representing erosion in a sampling unit using a summary statistic such as the local average and the variance.
However, summary statistics do not characterize local details. For example, they could be insensitive to the most eroded lands (i.e., tail probabilities), which are the focus of erosion management.
There are also parametric approaches \citep[e.g.,][]{gelf:05,duns:08} assuming the spatially-indexed distributions reside within a (often customized) parametric family where the dependency on covariates and locations is handled through parameters via hierarchy, yet it could be non-trivial to find a suitable simple family for, e.g., as demonstrated in \autoref{fig:sample.unit}, skewed and multi-modal local distributions.

Quantile regression \citep[e.g.,][]{koen:17} provides a valuable framework for modeling conditional distributions. However, it faces challenges such as quantile crossing (i.e., higher quantiles exceeding lower ones) that often require post-hoc adjustments \citep{he:97,reic:11,xu:23}. In this work, we instead focus on approaches that directly treat entire distributions as modeling objects, as developed in functional data analysis.

Modeling probability distributions as standalone objects is common in functional data analysis. 
Using functional variograms and Kriging, spatially distributed functional objects have been studied \citep{koko:17,zhan:21,mate:21}. However, these approaches operate within a Hilbert space, whereas the space of probability distributions even lacks a linear structure since there is no inherent addition operator. As a result, existing functional Kriging, being a linear predictor, may yield predictions that are not valid probability distributions.

To address this issue and accommodate the nonlinear constraints of distributions (positivity and unit integral), a typical procedure is to transform the distribution to a linear vector space for analysis, and then map the results back to the distributional space \cite[e.g.,][]{pete:16, hron:16}.
For instance, \citet{mena:14,mena:21} analyzed probability distributions from a compositional data perspective using the Bayes Hilbert space framework \citep{boog:14}, in which the addition of two distributions is defined through the product of their probability densities. However, it represents distributions in a linear space via the centered log-ratio transform, which requires all densities to be strictly positive over the entire domain of interest. This assumption makes the Bayes Hilbert space unsuitable because erosion distributions, especially their support, vary drastically across locations, leaving no common support on which the centered log-ratio can be applied.

Wasserstein space \citep{pana:20} has recently emerged as a promising framework, representing probability distributions in a linear tangent space. Similar ideas of embedding data into a tangential Hilbert space have been used for Kriging symmetric positive definite matrices \citep{pigo:16}. However, our focus on probability distributions involves a fundamentally different geometry.
Within the Wasserstein framework, \frechet regression can be applied to model distributional responses based on covariates \citep{pete:19-1,bhat:23,ghos:23,tuck:25}. However, as we later discovered in \autoref{sec:simulation}, existing \frechet regression approaches are suboptimal in our setting since spatial dependency is unaccounted for.
On the other hand, \cite{balz:20} also worked within the Wasserstein space, defining a variogram and ordinary Kriging of histograms to capture the spatial dependence of sensor data stream distributions. This approach, however, offers no straightforward way to incorporate covariates. In summary, existing approaches in the Wasserstein space address either covariates or spatial dependence, but not both aspects simultaneously. Moreover, operating directly on quantile functions or  histograms, they often incur high computational cost.

Our work applies Wasserstein space to model spatially correlated distributions while explicitly incorporating covariates. Building on this foundation, we now outline the framework proposed and summarize our contributions.

\subsection{Summary and Our Contributions}\label{sec:contribute}

We propose to treat such a local distribution as a random element in the Wasserstein space, which is then modeled utilizing the linearity of the associated tangent bundle.
We further represent the tangential element via basis expansion to ease computation, where the expansion coefficients capture the spatial structure.
Subsequently, the multivariate random fields of coefficients are analyzed via conventional geostatistical methods, allowing covariates to be incorporated.
For added flexibility, we generalize universal Kriging to allow the potential nonlinear effect of spatial covariates.
In the end, results for the multivariate random fields are brought back to the original Wasserstein space of distributions via an inverse transformation. In this way, our approach models spatially correlated distributions in the Wasserstein space while explicitly accounting for covariates, bridging the gap in existing methods based on the Wasserstein space.

Our work develops a regression framework for spatially distributed random distributions (e.g., erosion), explicitly accommodating spatial covariates with nonlinear effects.
It provides a unified model that enables simultaneous investigation of key quantities, such as local averages, standard deviations, and quantiles, without invoking separate model components that are otherwise necessary as in Kriging. 
Moreover, in contrast to parametric approaches, our model adapts to the underlying data patterns, making it highly attuned to real-world complexities.
Like Kriging, our model separates the covariate-driven trend from local stochastic dependencies, offering enhanced interpretability and flexibility.
This separation is well-grounded in the pseudo-Riemannian geometry of the Wasserstein space, ensuring both theoretical soundness and practical applicability.

In addition, our approach's adoption of basis expansion significantly reduces computational costs. For example, analysis for Shaanxi province (more than 205,000 $km^2$) over a 1km-by-1km grid was completed within 7 hours using 80 cores on a cluster consisting of multiple 8-core Intel E5-2640 v3 CPUs, demonstrating the model’s efficiency for large-scale spatial analysis.

The rest of our work is organized as follows. 
\autoref{sec:method} details the proposed method with additional mathematical background for Wasserstein space included in 
\suppref{Section S1}
of the Supplement \citep{erosion:supp}.
\autoref{sec:estimation} discusses the model estimation and spatial prediction. 
\autoref{sec:simulation} investigates numerical performance via simulations.
In \autoref{sec:ern} we revisit the motivating soil erosion example utilizing the proposed approach.

\section{Modeling Spatially Distributed Random Distributions in the Wasserstein Space}\label{sec:method}
Denote the Wasserstein space of distributions (of soil erosion values) supported in domain $\cbx T \subset \dsR$ by $
\waspace(\cbx T) = \{\mu \text{ is a probability measure on } \cbx T: \int_\cbx T x^2 d \mu(x) < \infty\}.
$
Let $\Lambda: \calS \times \Omega \to \waspace(\cbx T)$ be a distribution-valued random field over the spatial domain $\calS \subset \dsR^2$, and write $\Lambda_s=\Lambda(s,\cdot)$ as the random distribution at location $s \in \calS$.
In our application, $\Lambda$ is the spatial random field of local erosion distributions, and $\Lambda_s$ is the erosion distribution at location $s$. 
We propose a model for the $\waspace$-valued random field $\Lambda$, formulated as
\begin{equation}
    \Lambda_s 
    =
    \wassexp[\lambda_\fave]\left( X_s \right), 
    \text{ with } 
    X_s(t) = \sum_k  \eta_k(s) \vphi_k(t) \in L^2_{\lambda_\fave}
    \text{, where }
    \eta_k(s) = g_k(u(s)) + z_k(s). \label{mdl:x_working}
\end{equation}
Essentially, the distribution $\Lambda_s$ is expressed via the expansion coefficients $[\eta_k(s)]_k$, forming a multivariate random field over $s\in\calS$ driven by the covariates $u(s)$.
Here $X_s$ is the linearization of the distribution $\Lambda_s$ lying in the Hilbert space $L^2_{\lambda_\fave}$, where 
the baseline measure $\lambda_\fave$ serves as the reference and will be detailed in \autoref{sec:just}; 
and the exponential map $\wassexp[\lambda_\fave]$ bring the linearization back to the space of distributions.
The function $X_s$ is represented via an orthogonal basis $\vphi_1, \dots, \vphi_K$, and the basis coefficients $\eta_k(s)$ captures the spatial variation.
The spatial variation is further decomposed into the deterministic spatial trend captured by a function $g_k(u(s))$ based on known spatial covariates $u$. The local dependency is captured by the spatial correlation of a zero-mean random field $z_k$. For instance, we assume that the covariance follows a pre-specified cross-covariance model in the form of
\begin{equation}\label{mdl:x_sp_cross.cov}
    \cov \left(\eta_k(s), \eta_{k'}(s') \right) = \cov \left(z_k(s), z_{k'}(s') \right) 
    = 
    C_{kk'}(\norm{s - s'} |\ \vec\rho_{kk'})
\end{equation}
for any $s, s' \in \calS$, where $\norm{s - s'}$ is the distance between $s$ and $s'$; and for all $k, k' = 1, \dots, K$ the $C_{kk'}(\cdot |\ \vec\rho_{kk'}): \dsR^+ \to \dsR^+$ is a non-increasing function that tends to zero as the first argument increases and is parameterized by $\vec\rho_{kk'}$.

We next define $\lambda_\fave$ and motivate model \eqref{mdl:x_working}, which decomposes the response into a deterministic covariate effect and a stochastic component.

\subsection{Separating the Deterministic Trend}\label{sec:just}

\begin{figure}
    \centering
    \resizebox{0.75\textwidth}{!}{\tikzset{every picture/.style={line width=0.75pt}} 

\begin{tikzpicture}[x=0.75pt,y=0.75pt,yscale=-1,xscale=1]

\draw  [color={rgb, 255:red, 255; green, 255; blue, 255 }  ,draw opacity=1 ][fill={rgb, 255:red, 74; green, 144; blue, 226 }  ,fill opacity=0.2 ] (241.43,26.68) -- (639,129.52) -- (232.61,345.14) -- (88.42,94.4) -- cycle ;
\draw [color={rgb, 255:red, 0; green, 0; blue, 255 }  ,draw opacity=1 ]   (250.21,172.16) -- (421.29,182.08) ;
\draw [shift={(423.28,182.19)}, rotate = 183.32] [color={rgb, 255:red, 0; green, 0; blue, 255 }  ,draw opacity=1 ][line width=0.75]    (10.93,-3.29) .. controls (6.95,-1.4) and (3.31,-0.3) .. (0,0) .. controls (3.31,0.3) and (6.95,1.4) .. (10.93,3.29)   ;
\draw  [dash pattern={on 4.5pt off 4.5pt}]  (250.21,172.16) .. controls (339,176.14) and (409,246.14) .. (432.02,317.54) ;
\draw [shift={(432.02,317.54)}, rotate = 72.13] [color={rgb, 255:red, 0; green, 0; blue, 0 }  ][fill={rgb, 255:red, 0; green, 0; blue, 0 }  ][line width=0.75]      (0, 0) circle [x radius= 3.35, y radius= 3.35]   ;
\draw [shift={(250.21,172.16)}, rotate = 2.56] [color={rgb, 255:red, 0; green, 0; blue, 0 }  ][fill={rgb, 255:red, 0; green, 0; blue, 0 }  ][line width=0.75]      (0, 0) circle [x radius= 3.35, y radius= 3.35]   ;
\draw  [dash pattern={on 3.75pt off 7.5pt}]  (512.06,268) .. controls (464,209.34) and (394.89,150.83) .. (250.21,172.16) ;
\draw [shift={(250.21,172.16)}, rotate = 171.61] [color={rgb, 255:red, 0; green, 0; blue, 0 }  ][fill={rgb, 255:red, 0; green, 0; blue, 0 }  ][line width=0.75]      (0, 0) circle [x radius= 3.35, y radius= 3.35]   ;
\draw [shift={(512.06,268)}, rotate = 230.67] [color={rgb, 255:red, 0; green, 0; blue, 0 }  ][fill={rgb, 255:red, 0; green, 0; blue, 0 }  ][line width=0.75]      (0, 0) circle [x radius= 3.35, y radius= 3.35]   ;
\draw [color={rgb, 255:red, 0; green, 0; blue, 255 }  ,draw opacity=1 ]   (250.21,172.16) -- (446.39,141.12) ;
\draw [shift={(448.37,140.81)}, rotate = 171.01] [color={rgb, 255:red, 0; green, 0; blue, 255 }  ,draw opacity=1 ][line width=0.75]    (10.93,-3.29) .. controls (6.95,-1.4) and (3.31,-0.3) .. (0,0) .. controls (3.31,0.3) and (6.95,1.4) .. (10.93,3.29)   ;
\draw [color={rgb, 255:red, 0; green, 0; blue, 255 }  ,draw opacity=1 ][line width=0.75]    (250.21,172.16) -- (274.25,132.48) ;
\draw [shift={(275.29,130.77)}, rotate = 121.22] [color={rgb, 255:red, 0; green, 0; blue, 255 }  ,draw opacity=1 ][line width=0.75]    (10.93,-3.29) .. controls (6.95,-1.4) and (3.31,-0.3) .. (0,0) .. controls (3.31,0.3) and (6.95,1.4) .. (10.93,3.29)   ;
\draw    (18.14,328.83) .. controls (161.12,94.3) and (354.3,13.72) .. (618.93,11.21) ;

\draw (213.38,167.96) node [anchor=north west][inner sep=0.75pt]  [font=\Large]  {$\lambda _{\oplus }$};
\draw (83.9,302.76) node [anchor=north west][inner sep=0.75pt]  [font=\Large]  {$\waspace$};
\draw (211,35) node [anchor=north west][inner sep=0.75pt]  [font=\Large,color={rgb, 255:red, 0; green, 0; blue, 255 }  ,opacity=1 ]  {$T_{\lambda _{\oplus }}\waspace$};
\draw (452.6,175.87) node [anchor=north west][inner sep=0.75pt]  [font=\Large,color={rgb, 255:red, 0; green, 0; blue, 255 }  ,opacity=1 ]  {$G_{s} :=\log_{\lambda _{\oplus }} \lambda _{s}{} \ $};
\draw (387.9,306.33) node [anchor=north west][inner sep=0.75pt]  [font=\Large]  {$\lambda _{s}$};
\draw (475.16,129.46) node [anchor=north west][inner sep=0.75pt]  [font=\Large,color={rgb, 255:red, 0; green, 0; blue, 255 }  ,opacity=1 ]  {$X_{s} :=\log_{\lambda _{s}} \Lambda _{s}$};
\draw (266.54,105.25) node [anchor=north west][inner sep=0.75pt]  [font=\Large,color={rgb, 255:red, 0; green, 0; blue, 255 }  ,opacity=1 ]  {$\zeta _{s}$};
\draw (486.4,279.96) node [anchor=north west][inner sep=0.75pt]  [font=\Large]  {$\Lambda _{s} =\exp_{\lambda _{\oplus }} X_{s}$};

\end{tikzpicture}}
    \caption{
        A visualization for the proposed model, where the quantities are
        $\Lambda_s$: Random distribution at location $s$ in $\waspace$;
        $\lambda_\fave$ and $\lambda_s$: the reference measure and local expectation in $\waspace$;
        $T_{\lambda_\fave} \waspace$ (the plane): tangent space at $\lambda_\fave$;
        $G_s = \log_{\lambda_\fave} \lambda_s$, logarithm of the local expectation on $T_{\lambda_\fave} \waspace$;
        $X_s$, the logarithm of $\Lambda_s$ on $T_{\lambda_\fave} \waspace$, further decomposed as the sum of $G_s$ and $\zeta_s$.
    }\label{fig:wass_tangent_simple}
\end{figure}
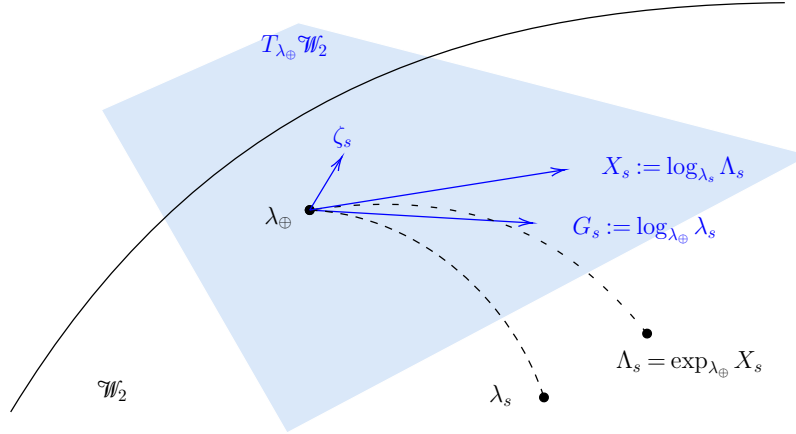

Model \eqref{mdl:x_working} exploits the linear structure of the tangent space of $\waspace$ at the reference measure $\lambda_\fave$ as illustrated in \autoref{fig:wass_tangent_simple}, to be defined shortly. 
Denoted as $T_{\lambda_\fave} \waspace$, the tangent space is a closed convex subset of $L^2_{\lambda_\fave}$, the space of squared-integrable functions with respect to $\lambda_\fave$ \citep[e.g.,][]{bigo:17}.
The logarithm map $\log_{\lambda_\fave}$ maps the probability distributions from $\waspace$ to the tangent space. 
We refer to \suppref{Section S1}
of the Supplement \citep{erosion:supp} for details of the pseudo-Riemannian structure of the $\waspace$ and the logrithm/exponential maps.

Subsequently, the random distribution $\Lambda_s$ is transformed to a random element $X_s = \log_{\lambda_\fave}(\Lambda_s)$ within the Hilbert space $L^2_{\lambda_\fave}$.
Similarly, we transform the local expectation of $\Lambda_s$, denoted as $\lambda_s$, that captures the deterministic \emph{trend} to another squared-integrable function $G_s \defas \log_{\lambda_\fave}(\lambda_s) \in L^2_{\lambda_\fave}$, while the remaining random element $\zeta_s = X_s - G_s$ is responsible for spatial dependency.
In the end, projecting $G_s$ and $\zeta_s$ to an orthogonal basis leads to model \eqref{mdl:x_working}.

More precisely, given the random measure $\Lambda_s$  at location $s$, write $\lambda_s$ as the local \frechet mean (over the probability space) and define the reference as a spatial \frechet mean (further, over the spatial domain) $\lambda_\fave$:
\begin{align}
    \lambda_s = \argmin_{\gamma \in \waspace} \Exp \wassdsq{\gamma}{\Lambda_s},
    \quad
    \label{def:sp_frechet_mean}
    \lambda_\fave = 
    \argmin_{\gamma \in \waspace} \int \Exp \wassdsq{\gamma}{\Lambda_s} d \cbx n(s),
\end{align}
where $\wassdsq{\cdot}{\cdot}: \waspace \times \waspace \to \dsR^+$ is the 2-Wasserstein distance, and $d \cbx n(s)$ is the sampling process of reference points (e.g., spatial distribution of the surveyed sampling units).

The local \frechet mean $\lambda_s$ captures the deterministic spatial \emph{trend} via
$
G_s \defas \wasslog[\lambda_\fave] \left( \lambda_s \right)
\in 
L^2_{\lambda_\fave}.
$
The spatial \frechet mean $\lambda_\fave$ centers the response with $\int_\calS G_s d \cbx n (s) = 0$.
Denoting the remaining zero-mean random component as
$
\zeta_s \defas \wasslog[\lambda_\fave] \Lambda_s - G_s
$,
the $\waspace$-valued random field $\Lambda_s$ is mapped to a $L^2_{\lambda_\fave}$-valued random field $\cbrk{X_s: s \in \calS}$ following
\begin{equation}\label{mdl:x}
    X_s(t) = G(u(s), t) + \zeta_s(t) \text{ for } t \in \cbx T,
\end{equation}
for some trend component $G^{}: \dsR^p \times \dsR \to \dsR$ determined by spatially varying covariates $u: \calS \to \dsR^p$.
The exponential map translates them back to $\waspace$ via the exponential map:
$
\wassexp[\lambda_\fave] X_s = \wassexp[\lambda_\fave] \left( G_s + \zeta_s \right),
$
constituting a model for $\Lambda_s$.

A more intuitive perspective utilizes quantile functions, to which the logarithm and exponential maps are closely related under the univariate setup. 
The logarithm map is $X_s := \log_{\lambda_\fave} \Lambda_s = \qf[\Lambda_s] \circ \cdf[\lambda_\fave] - \id$, where $\qf$ and $\cdf$ are corresponding quantile and distribution functions respectively. 
Suppose the $\cdf[\lambda_\fave]$ is continuous and strictly increasing, we have \begin{equation}\label{eq:x_simple_qf}
    X_s(t) = \qf[\Lambda_s](q) - \qf[\lambda_\fave](q)
\end{equation}
with $t = \qf[\lambda_\fave](q)$ for $q \in (0, 1)$.
In other words, the linearization of $\Lambda_s$ is the warped quantile functions. 
Subsequently, \frechet mean in $\waspace$ is identified by expectation of quantile functions \citep[e.g.,][theorem 3.2.11]{pana:20}:
$\qf[\lambda_s] = \Exp \qf[\Lambda_s]$ and $\qf[\lambda_\fave] = \int \Exp\qf[\Lambda_s] d\cbx n(s)$.
This implies $\lambda_\fave$ is also the \frechet mean of $\cbrk{\lambda_s, s \in \calS}$.
Under the quantile representation, the presented problem is effectively a functional regression problem upon the quantile functions with the addition of spatial dependency.

\subsection{Dimension Reduction via Basis Expansion}

After the preceding decomposition, it is still difficult to tackle $G_s$ and $\zeta_s$ as dependent functional objects. 
Further simplification is possible exploiting the fact that the tangent space $T_{\lambda_\fave}\waspace$ is a closed convex subset of $L^2_{\lambda_\fave}$.
We adopt basis expansion as a dimension reduction tool to separate the functional variation (in $t$) from spatial dependency (in $s$).

Consider an orthonormal basis $\vphi_1, \vphi_2, \dots$ in $L^2_{\lambda_\fave}$. By basis expansion, a $K$-dimensional representation for the tangential element is
$
X_s(t) \approx X^{}_{s, K} (t) = \sum_{k=1}^K \eta_k (s) \vphi^{}_k(t)$ for $t \in \cbx T$,
where $\eta_k(s) = \innerL[\lambda_\fave]{X^{}_s}{\vphi^{}_k}$ with the inner product taken with respect to $\lambda_\fave$. 
Thus the orthonormal basis carries the functional variation, while the random fields $\eta_1, \dots, \eta_K$ carry spatial structure over $s \in \calS$.
Under model \eqref{mdl:x}, the random and deterministic random fields of respective basis coefficients are
$
z_{k}(s) = \innerL[\lambda_\fave]{ \zeta_s^{} }{\vphi_k^{}}, 
g_{k} (u)  = \innerL[\lambda_\fave]{G^{}(u, \cdot)}{\vphi_k^{}}, 
$
$k=1,\dots,K$.
In sum, we have
\begin{equation}\label{mdl:x_basis_coef}
    \eta_k(s) = g_k \circ u(s) + z_k(s), \text{ for } k = 1, \dots, K,
\end{equation}
where $\eta_k$, $g_k$, and $z_k$ are scalar random fields that are easier to work with.

Moreover, by orthonormality,
$
\wassdsq{\Lambda_{s_1}}{\Lambda_{s_2}} = \normL[\lambda_\fave]{X^{}_{s_1} - X^{}_{s_2}}^2 =
\sum_{k=1}^K \left(\eta_k(s_1) - \eta_k(s_2)\right)^2
+ \sum_{k>K} \left(\eta_k(s_1) - \eta_k(s_2)\right)^2
$
where the second summation is small if $K$ is appropriately large,
so that the Euclidean distance between the first $K$ expansion coefficients approximates the Wasserstein distance between the probability distributions.
Thus, model \eqref{mdl:x} of the $L^2_{\lambda_\fave}$-valued random field $\Lambda$ translates to model \eqref{mdl:x_basis_coef} on multivariate random field \eqref{mdl:x_basis_coef} while preserving the metric. 
In combine, we arrive at the proposed model \eqref{mdl:x_working}--\eqref{mdl:x_sp_cross.cov} for the distribution-valued random field.

\section{Estimation and Spatial Prediction}\label{sec:estimation}

\subsection{Estimating the Fixed Components} \label{ssec:fixedComponents}
To provide an overview of how we estimate the fixed components in model \eqref{mdl:x_working}--\eqref{mdl:x_sp_cross.cov}, we start by expressing distributional objects using the corresponding sample quantile functions. 
Next, we estimate the spatial \frechet mean \eqref{def:sp_frechet_mean} by also expressing it as a quantile function.
The observed local distributions are then mapped to tangential elements which are square-integrable trajectories. 
After specifying an orthonormal basis, we compute the empirical coefficients at observed locations, which are used to estimate parameters in the spatial covariance model \eqref{mdl:x_sp_cross.cov}, 
as well as the trend component (the $g_k (u(s))$ in \eqref{mdl:x_working}).

In detail, at any location $s_i$ we draw the empirical distribution from discrete observations $Y_{i1}, \dots, Y_{i, N_i}$ (soil erosion values) in accordance with the quantile functions $\empcdf[i]$ and $\empqf[i]$, respectively, which are regarded as ``noisy'' observation of the underlying distribution $\Lambda_{s_i}$, for $i=1,\dots,n$. 
Represented as a quantile function, the empirical spatial \frechet mean is defined through the mean of the sample quantile functions, obtaining 
$
\empqf[\fave](q) = n^{-1} \sum_{i=1}^n \empqf[i](q) \text{ on $q \in (0, 1)$, }
$
which solves the empirical version of \eqref{def:sp_frechet_mean}: minimizing
$
n^{-1} \sum_{i=1}^n \int_0^1 \left(\qf[\gamma](q) - \empqf[i](q)\right)^2 dq
$
over all possible quantile functions $\qf[\gamma]$.
The measure associated with $\empqf[\fave](q)$ is denoted as $\emp \lambda_\fave$, which estimates the spatial \frechet mean $\lambda_\fave$. 
The empirical tangential element representing local distribution $\Lambda_{s_i}$ is
$
\emp X_i = \empqf[i] \circ \empcdf[\fave] - \id
$
,
where $\empcdf[\fave]$ is the cumulative distribution function corresponding to $\empqf[\fave]$.

To obtain the most parsimonious representation of the distribution-valued spatial process $\Lambda_s$, we perform the eigendecomposition of the tangential elements $\emp X_1, \dots, \emp X_n$ and adopt the first $K$ eigenfunctions as the basis functions.
Note that such a decomposition is taken under the baseline measure $\emp \lambda_\fave \in \waspace$ so the integral is with respect to $\emp\lambda_\fave$.
Since the tangential elements have mean $\smplaveinline[n]{i} \emp X_{s_i} = 0$, we eigen-decompose the erosion covariance function
$
\emp{\calG}(t_1, t_2) = \smplaveinline[n]{i} \emp X_{s_i}(t_1) \emp X_{s_i}(t_2) \text{ for } t_1, t_2 \in \cbx T
$
and obtain eigenvalues $\emp\sigma_1 \geq \emp\sigma_2 \geq \cdots$ and eigenfunctions $\emp \vphi_k$ as the solutions to the eigen-problem
\begin{equation}\label{eq:cov_surface_eig_problem}
    \emp\sigma_k \emp \vphi^{}_k(t_0) = \int \emp \calG(t_0, t) \emp \vphi^{}_k(t) d\emp \lambda_\fave(t) \text{ for all $t_0 \in \cbx T$, }
\end{equation}
where the eigenfunctions $\emp \vphi_1, \emp \vphi_2, \dots$ constitute an orthonormal basis for $L^2_{\emp\lambda_\fave}$, referred to as the eigenbasis.
Under the quantile representation \eqref{eq:x_simple_qf} and change of measure, this is effectively functional principal component analysis of quantile functions with respect to the Lebesgue measure.

In sum, the tangential elements are expressed in the eigenbasis as 
$
\emp X_{s_i}(t) \approx \sum_{k = 1}^K \emp\eta_k(s_i) \emp\vphi_k(t)
$
with the basis coefficients 
$
\emp \eta_k(s_i) = \innerL[\emp \lambda_\fave]{\emp X_{s_i}}{\emp \vphi_k}
$.
We then use the empirical coefficients $\emp \eta_k$ as estimates of the scalar random fields $\eta_k$. 

\subsection{Spatial Prediction} \label{sec:krigingPred}

To predict at a new location $s_0$, 
it suffices to predict $[\eta_k(s_0)]_{k=1}^K$ 
given observed $\Lambda_{s_1}, \dots \Lambda_{s_n}$ at locations $s_1, \dots, s_n \in \calS$. 
For notational simplicity, we suppose $\vec \eta_k = \left(\eta_{k}(s_1), \dots, \eta_{k}(s_n) \right)\tpose$ is available and later plug in their empirical version $\emp\eta$. 
The best linear unbiased predictor (BLUP) for $\eta_k(s_0)$ is 
\begin{equation}\label{eq:blup_krige}
    \text{BLUP} \left(\eta_{k}(s_0) | \vec \eta_k \right)
    =
    \Exp \eta_{k}(s_0) + \cov(\eta_{k}(s_0), \vec \eta_k) (\cov \vec \eta_k )\inv (\vec \eta_k - \Exp \vec \eta_k ) ,
\end{equation}
where the trend model is $\Exp \eta_{k}  = g_k\circ u$ and the covariance is modeled in \eqref{mdl:x_sp_cross.cov}.

The spatial prediction procedure utilizing a covariance model and BLUP is usually referred to as Kriging in geostatistics \citep{cres:93, stei:99, gelf:10}. 
While the $k$th basis coefficient $\eta_k(s)$ could in principle be estimated through co-Kriging which aggregates information across all coefficients $\eta_{k'}$, $k'=1,\dots,K$, we find in preliminary investigation that coefficients $\eta_{k'},\, k'\ne k$ provide only limited information about $\eta_k$ at the cost of substantially heavier computation (see 
\suppref{Subsection S8.2}
of the Supplement \citep{erosion:supp}). We thus predict $\eta_k$ using only spatial observations from the same coefficient across different locations as in \eqref{eq:blup_krige}.

To account for the nonlinear covariate effects, we modify the universal Kriging approach, replacing the usual linear trend model by local polynomial regression.
Denote
\begin{align}
    \vec \eta_k = \left(\eta_k(s_1), \dots, \eta_k(s_n)\right)\tpose, 
    &\quad 
    \vec z_k = \left(z_k(s_1), \dots, z_k(s_n)\right)\tpose, \nonumber\\ 
    \vec v_{0k} = \cov\left(\vec \eta_k, \eta_k(s_0)\right) = \cov\left(\vec z_k, z_k(s_0)\right),
    &\quad 
    \mat{\Sigma}_k = \cov\left(\vec \eta_k\right) = \cov\left(\vec z_k\right), \label{eq:eta_sp_cov}
\end{align}
where $\vec v_{0k}$ and $\mat{\Sigma}_k$ are determined by spatial location and covariance model \eqref{mdl:x_sp_cross.cov}.

By the Kriging formula, the predictor for $k$th coefficient at $s_0$ is
\begin{align}
    \tl \eta_k(s_0) = \tl g_k\circ u(s_0) + \tl z_k(s_0) &\defas \vec l_{0k}\tpose \vec \eta_k + \vec v_{0k}\tpose \mat{\Sigma}_k\inv(\mat{I} - \mat{L}_k)\vec \eta_k, \label{eq:blup_estiE}
\end{align}
where $\tl g_k \circ u(s_0) = \vec l_{0k}\tpose \vec \eta_k$ is the local polynomial estimate of the trend component at $s_0$, $\vec l_{0k}$ is the vector of coefficients from the local polynomial regression determined by covariates $u$,
and $\tl z_k(s_0)=\vec v_{0k}\tpose \mat{\Sigma}_k\inv(\mat{I} - \mat{L}_k)\vec \eta_k$ is the simple Kriging predictor. 
Here, we plug in the local polynomial regression estimator $\mat{L}_k \vec \eta_k$ to tackle the trend component $\Exp\vec\eta_k$ at the observed locations, where $\mat{L}_k$ is solely determined by the known covariate values at $s_0, \dots, s_n$ and detailed in 
\suppref{Section S3}
of the Supplement \citep{erosion:supp}.

In the end, pooling coefficient-wise predictions together gives the predicted measure at $s_0$ as
$
\tl \Lambda_{s_0} = \wassexp[\lambda_\fave] \left(
\sum_{k=1}^K \tl \eta_{k}(s_0) \vphi_k
\right)
$.
To predict any given functional $\functional: \waspace \to \dsR$ (e.g., the mean) of the probability measure, we define a multivariate function
\begin{equation}\label{eq:score2functional}
    \mathsf{f}: \dsR^K \rightarrow \dsR,\quad \mathsf{f}(\eta_1, \dots, \eta_K) = \functional \circ \wassexp[\lambda_\fave] \left( \sum_{k=1}^K \eta_k \vphi_k \right),
\end{equation}
mapping from the basis coefficients to the functional value. The predicted functional value is then $\mathsf{f}(\tl\eta_1(s_0), \dots, \tl\eta_K(s_0))$.

\subsection{Uncertainty Quantification}\label{sec:uncertainty}

We use the variance of prediction error
to assess the uncertainty of the BLUP in \eqref{eq:blup_estiE}, which is characterized by the following proposition. It incorporates uncertainty in both the local polynomial estimation of the trend component $g_k \circ u$ and the random component $z_k$.
\begin{ps}\label{prop:var}
    Under \eqref{eq:blup_estiE},
    \begin{align*}
        \var \left(z_k(s_0) - \tl z_k(s_0)\right) &= \var z_k(s_0) - \vec v_{0k}\tpose \mat{\Sigma}_k\inv \vec v_{0k}, \\
        \var(\eta_k(s_0) - \tl \eta_k(s_0))
        &= 
        \var\left(z_k(s_0) - \tl z_k(s_0)\right) \\
        + \vec l_{0k}\tpose \mat{\Sigma}_k \vec l_{0k}  
        &+ \vec v_{0k}\tpose \mat{\Sigma}_k\inv \mat{L}_k \mat{\Sigma}_k \mat{L}_k\tpose \mat{\Sigma}_k\inv \vec v_{0k}
        - 2 \vec l_{0k}\tpose \mat{\Sigma}_k \mat{L}_k\tpose \mat{\Sigma}_k\inv \vec v_{0k}. 
    \end{align*}
\end{ps}
Under a Gaussian assumption, the uncertainty of the predicted coefficients $\tl\eta_k$ propagates to that of predicted local distributions and also any functionals.
Assuming $z_k$ and hence $\eta_k$ are Gaussian, we can generate coefficients $\eta^*_{1k}, \dots \eta^*_{Jk}$ from a Gaussian distribution with mean $\tl\eta_k(s_0)$ and variance $\var\left(\eta_k(s_0) - \tl \eta_k(s_0)\right)$, $k = 1, \dots, K$ for a sufficiently large $J$ Monte Carlo repeats. Then an equal-tail $(1-\alpha)$ prediction interval for $\functional(\Lambda_{s_0})$ is given by the $(\alpha/2, 1-\alpha/2)$-sample quantiles of $\cbrk{\mathsf{f}(\eta^*_{j1}, \dots, \eta^*_{jK}): j = 1, \dots, J}$. 

While \autoref{prop:var} is still valid without the Gaussianity assumption, the prediction interval based on the generated Gaussian sample may fail. One alternative is to assume the random component $z_k$ is strictly stationary, so that we approach the marginal distribution of $z_k(s_0)$ via the observations/estimations pooled over the spatial domain $\cal Z = \cbrk{z_k(s_1), \dots, z_k(s_n)}$. We then draw the Monte Carlo sample $\cbrk{\eta^*_{1k}, \dots \eta^*_{Jk}}$ with $\eta^*_{jk} = \tl\eta_k(s_0) + z^*_{jk}$ for $j = 1, \dots, J$, where $z^*_{jk}$ is re-sampled from $\cal Z$ and proceed similarly. 

\subsection{Plug-in Estimation}

For simplicity, we fit the spatial covariance model \eqref{mdl:x_sp_cross.cov} utilizing the sample variogram of the residuals of local polynomial regression of $\emp{\eta}_k(s_i)$ on $u(s_i)$, $i = 1, \dots, n$, similar to that of universal Kriging.
Model \eqref{mdl:x_sp_cross.cov} provides a plug-in estimator for the covariance matrix ${\mat{\Sigma}}_k$. The BLUE \eqref{eq:blup_estiE} is then computed as $
\h\eta_k(s_0) = \vec l_{0k}\tpose \emp{\vec \eta}_k + \h{\vec v}_{0k} \tpose \h{\mat{\Sigma}}_k\inv(\mat{I} - \mat{L}_k)\emp{\vec \eta}_k.
$
The predicted measure at $s_0$ is obtained as
$
\h \Lambda_{s_0} = \wassexp[\emp \lambda_\fave] \left(
\sum_{k=1}^K \h \eta_{k}(s_0) \emp \vphi_k
\right)
$
utilizing the exponential map.
For prediction uncertainty, we plug the corresponding estimated quantities into \autoref{prop:var} and follow \autoref{sec:uncertainty}.

By the definition of logarithm map, any tangential element $X_s$ in the image of $\log_{\lambda_\fave}$ must satisfy that $X_s + \id = \qf[\Lambda_s] \circ \cdf[\lambda_\fave]$ is a non-decreasing function.
Hence, to be properly mapped back to $\waspace$, the truncated representation $X_{s, K}(t) + t = \sum_{k=1}^K \eta_k(s) \vphi_k(t) + t$ needs to be non-decreasing over $t \in \cbx T$.
To enforce such monotonicity, we apply an \emph{ad hoc} monotonization, defined for a function $f$ as
$
\upup{f} (x) = \max\left( f(x), \ \sup_{t \leq x} f(t)\right).
$
The monotonized predicted measure is obtained as
$
\h \Lambda_{s_0} 
\approx \upup{\left(\sum_{k=1}^K \h \eta_{k}(s_0) \emp \vphi_k + \id\right)}\pushforward \emp\lambda_\fave
$ where $f\pushforward \mu=\mu\circ f^{-1}$ denotes the pushforward of measure $\mu$ by $f$.

The number $K$ of bases and the form of the spatial covariance function $C_{kk'}$ in (\ref{mdl:x_sp_cross.cov}) must be specified.
We propose to choose $K$ using the fraction of variation explained (FVE) criterion \citep[see e.g.,][]{pete:22} and select the smallest number of components such that $\text{FVE}(K)\coloneqq \sum_{k = 1}^K \emp\sigma_k / \sum_{j=1}^\infty \emp\sigma_j$ exceeds 95\%, recalling that $\emp\sigma_1, \dots \emp\sigma_K, \dots$ are the eigenvalues defined in \eqref{eq:cov_surface_eig_problem}. 
The spatial covariance function is modeled using the \matern covariance function with smoothness, range, sill, and nugget effect parameters estimated via sample variogram (See 
(\suppref{S6.1}) 
of the Supplement \citep{erosion:supp} for our parameterization and also \cite{stei:99}.)

\section{Simulation}\label{sec:simulation}

In the following simulation study, we illustrate the proposed method under a scenario resembling the erosion data yet simplified. We outline the results here and refer to 
\suppref{Section S6} 
of the Supplement \citep{erosion:supp} for details, additional performance assessment, and more simulation settings that further highlight the flexibility of the proposed method.

At any given $s\in\calS$ over the spatial domain $\calS = [0, 10] \times [0, 10]$, we drew samples $Q_{s, 1}, \dots, Q_{s, N_s}$ from a zero-hurdled log-mixture distribution. For $i = 1, \dots, N_s$, we have $Q_{s, i} = 0$ with probability $p_0(s)$ and $Q_{s, i} = \exp(Y_{s, i})$ with probability $1 - p_0(s)$, where $Y_{s, i}$ follows a Gaussian mixture of $N(\theta_1(s), 1)$ and $N(0, \theta_2(s)^2)$ with equal probability.
The parameters $p_0, \theta_1$, and $\theta_2$ of the local distribution are defined by
$
p_0(s) = 1 - \Phi( (\sin(u_0(s)) + z_1(s)) / 3 + 3/4),
$
$
\theta_1(s) = - 2 + 5 u_1(s) + z_1(s)
$, and $
\theta_2(s) = \exp\left(
-0.22 + 0.12 u_2(s) + z_2(s)
\right)
$,
where $\Phi$ is the distribution function of the standard normal distribution.
The $u_k: \calS \to \dsR$, $k = 0, 1, 2$ are the known spatial features used as covariates in modeling, while $z_k: \calS \to \dsR$, $k=1,2$ are two latent mean-zero Gaussian random fields. 
The spatial dependency for $z_1, z_2$ is specified by the \matern covariance function with range parameter $\rho = 1.5$, smoothness parameter $v = 1$, and some nugget effect that accounts for 5\% of variation.

In summary, the observations include the zero probability $p_0$, covariates $u_1, u_2$ and $n$ samples (independent conditioning on the $p_0, \theta_1, \theta_2$) of discrete observations $Q_{i1}, \dots, Q_{i, N_{i}}$ 
at locations $s_1, \dots, s_n \in \calS$. 
We randomly sampled $n = 500$ locations uniformly over the region $\calS$, and the number of observations in each local sample $N_i$, $i=1,\dots,n$ are determined by $p_0$ and an independent Poisson distribution so that on average there were $300$ positive observations per location.

We treat the local distribution in two parts: the point mass at zero and the distribution of the positive part. The point mass was modeled via $p_0 = \logit\inv(\eta_0)$ where the random field $\eta_0$ is tackled through the proposed generalized universal Kriging (\autoref{sec:krigingPred}).
Similar to the erosion data (\autoref{fig:sample.unit}), the remaining log-mixture distribution (that of $\exp Y$) is heavily right-skewed. Thus, we applied a log-scaling and worked on $Y$ directly. The predicted distribution function was transformed and mixed with the predicted zero-probability to reconstruct a prediction of the distribution in the original scale (i.e., that of $Q$). In particular, while the underlying true $\theta_1$ and $\theta_2$ each only relies on univariate feature $u_1$ and $u_2$ respectively, the proposed method was unaware of such truth and utilized both features while fitting.

The prediction performance is assessed in three aspects: 
(i) that of the local parameter via the generalized universal Kriging;
(ii) that of the local log-mixture distributions in log-scale;
and (iii) that of the local distributions after being mixed with predicted zero-probability in the original scale.

\autoref{tbl:simu_par_err} summarizes our findings for aspect (i). The proposed generalization significantly reduced prediction error compared to universal Kriging when the covariate effect is nonlinear (as in the case of $p_0$); while being comparable when the covariate effect is linear (as in the case of $\theta_1$ and $\theta_2$).

\begin{figure}
    \centering
    \includegraphics[width=0.9\textwidth]{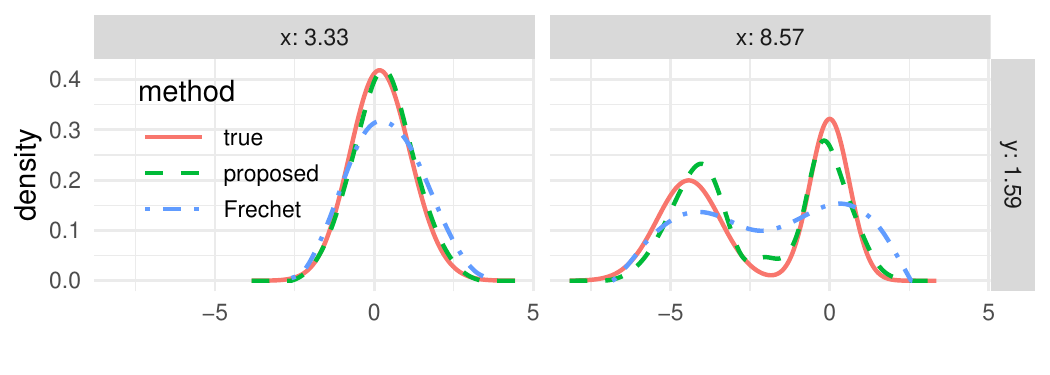}
    \caption{True and the predicted densities using a local \frechet regression and the proposed method. Each panel corresponds to a different spacial location with coordinate $(x, y)$. }
    \label{fig:simu_pdf}
\end{figure}

For aspect (ii), the proposed method was compared against parametric models utilizing either the normal family or the correct log-mixture family. Maximum likelihood estimation (MLE) was used to estimate the corresponding local parameters $(\theta_1, \theta_2)$ at the sampled locations. The proposed generalized universal Kriging methods were then applied to predict the local parameters, which were subsequently plugged into the parametric distribution model for predicting the underlying measure at any query location. 
For a fair comparison, we used $K = 2$ components in the proposed method to match the number of parameters in the parametric approaches.
Moreover, local \frechet regression \citep{pete:19-1}, which also utilizes Wasserstein geometry to model the local distributional objects as a function of location coordinates, was also included.
Prediction performance is assessed by the mean Wasserstein error (MWE), defined as 
\begin{equation}\label{def:mwe}
    \text{MWE} = 
    \int_{\calS} 
    \left( \int_0^1 \left( F_s\ginv(q) - \h F_s\ginv(q) \right)^2 dq  \right)^{1/2}
    d\cbx n (s),
\end{equation}
where $F_s\ginv$ and $\h F_s\ginv$ are the quantile function and its prediction at location $s$, and $\cbx n$ is the normalized intensity for spatial sampling process. Here $d \cbx n(s) = ds$ is the uniform sampling scheme.
The first row of \autoref{tbl:simu_cdf_err} shows that the proposed method outperformed all but the correctly specified parametric model in terms of distribution prediction in the log-scale (that of $Y$). \autoref{fig:simu_pdf} illustrates the true and some predicted densities at two locations. The normal model (not included in the figure) cannot capture the bi-modality, explaining its inflated errors. The \frechet regression neglected spatial dependency even though spatial coordinates were supplied as covariates, causing an even worse performance.

Lastly, for aspect (iii), we transformed the log-scale prediction (of $Y$) back to the original scale then mixed it with the predicted $p_0$ to reconstruct a predicted distribution of $Q$. The second row of \autoref{tbl:simu_cdf_err} lists the MWE to showcase the capability of the overall procedure: it outperformed all other approaches except the correctly specified parametric model.

In summary, the proposed procedure has been validated and outperforms existing approaches. Its performance is comparable to a correctly specified parametric MLE, which is usually challenging to obtain in practice. This simulation, whose data mimics the soil erosion data, is a prelude to the more extensive real data modeling.

\begin{table}
    \caption{Mean absolute error (average, minimum, and maximum over repeats) in parameter prediction, where the spatial prediction follows universal Kriging (linear trend) or the proposed generalized approach (nonlinear trend), utilizing parameters under the correct parametric model.}\label{tbl:simu_par_err}
    \centering
    \begin{tabular}[t]{l|rrr}
        \toprule
        & $p_0$ & $\theta_1$ & $\theta_2$ \\
        \midrule
        \cellcolor{gray!10}{linear trend} & \cellcolor{gray!10}{0.06 (0.05, 0.08)} & \cellcolor{gray!10}{0.31 (0.30, 0.33)} & \cellcolor{gray!10}{0.12 (0.08, 0.18)}\\
        nonlinear trend & 0.04 (0.03, 0.06) & 0.31 (0.30, 0.33) & 0.12 (0.09, 0.18)\\
        \bottomrule
    \end{tabular}
\end{table}

\begin{table}
    \caption{
        Errors in the predicted distributions. The correct and normal models fitted MLE within the correct or a normal distribution family respectively. The \frechet regression utilized the location coordinates as covariates.
        There were 30 repeated experiments for each simulation setup, where for each repeat we computed mean Wasserstein errors (MWE) of the predicted distributions under log or the original scale after mixed with zero-probability $p_0$. The averages (and the corresponding minimum and maximum) of the errors among the repeats are listed.
    }\label{tbl:simu_cdf_err}
    \centering
    \begin{tabular}[t]{l|cccc}
        \toprule
        Scale & correct model & proposed & normal model & Frechet regression\\
        \midrule
        \cellcolor{gray!10}{log} & \cellcolor{gray!10}{0.29 (0.28, 0.31)} & \cellcolor{gray!10}{0.39 (0.35, 0.46)} & \cellcolor{gray!10}{0.56 (0.44, 0.72)} & \cellcolor{gray!10}{1.07 (0.96, 1.34)}\\
        original with $p_0$ & 5.09 (0.96, 16.60) & 5.92 (1.07, 25.80) & 35.70 (2.31, 324.00) & 12.10 (1.94, 33.70)\\
        \bottomrule
    \end{tabular}
\end{table}

\section{Soil Erosion Modeling}\label{sec:ern}

\begin{figure}
    \centering
    \includegraphics[width=1\textwidth]{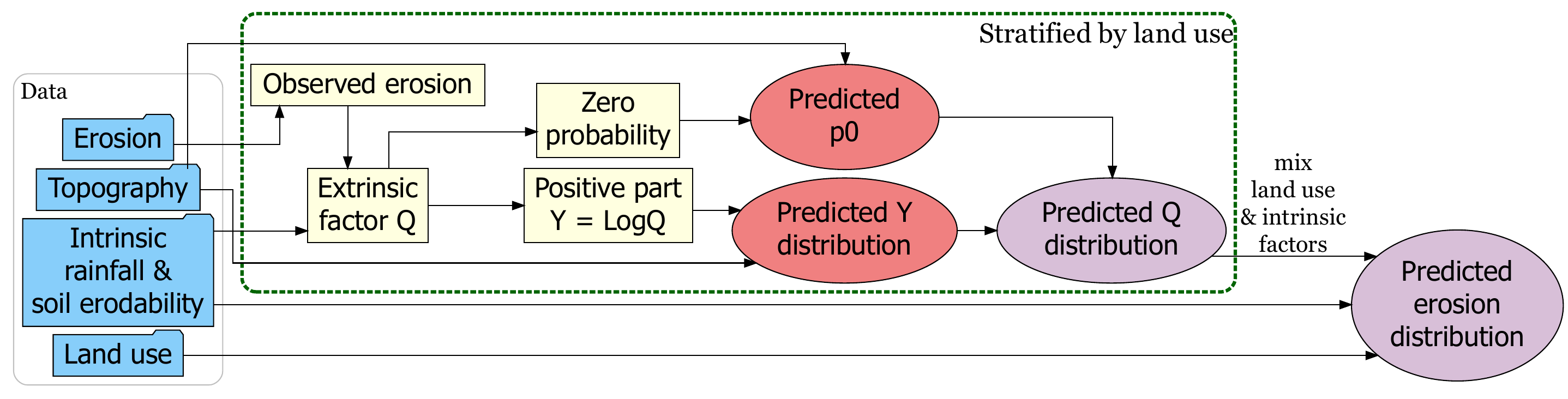}
    \caption{Flow chart for pre-processing, modeling, and prediction of erosion.}
    \label{fig:ern_flow}
\end{figure}

We now apply the proposed method to tackle the Shaanxi soil erosion data introduced in \autoref{sec:data}. Our modeling target is the local erosion distribution at any given location.
Since the sampling unit (small watersheds) is relatively tiny compared to the entire province, we use the coordinate $s$ of its geometric center to denote its location. Thus, the spatial distance between surveyed locations is calculated by that between their geometric centers (i.e., $\norm{s - s'}$ in \eqref{mdl:x_sp_cross.cov}). Further write $\ttv{A}_s$ as the random variable associated with the local erosion distribution. 
The empirical quantile function of detailed local erosion values at a surveyed sampling unit $s$ (at $10\times10$-meters resolution) was used as observation of the underlying truth.
We next detail a few additional model components imposed to accommodate the added complexity of offsets, conditioning, point mass, and skewness that present in the real data, for which \autoref{fig:ern_flow} provides a visualization.

The rainfall and soil erodibility components (\ttv{R} and \ttv{K}) are determined by climate and soil properties, which we refer to as \emph{intrinsic} factors.
Rewrite \eqref{def:soil_loss} as 
\begin{equation}\label{mdl:ern_rk_prod}
    \ttv{A}_s = \ttv{R}_s \times \ttv{K}_s \times Q_s,
\end{equation}
where
$Q_s = \ttv{L}_s \times \ttv{S}_s \times \ttv{B}_s \times \ttv{E}_s \times \ttv{T}_s$.We refer to $Q_s$ as the extrinsic factor, reflecting slope, vegetation/cover, and conservation practices, which are subject to management.
The intrinsic factors are readily available over the entire province from comprehensive meteorological data and soil survey data, leaving the extrinsic factor $Q_s$ as the key quantity for modeling. Here $Q_s$ and $(\ttv{R}_s, \ttv{K}_s)$ are assumed mutually independent, which is validated in \suppref{Subsection S8.1}. 

As illustrated in \autoref{fig:sample.unit}, erosion distribution can vary drastically for different land types even within the same sampling unit.
Rather than incorporating land type as a categorical variable directly into the local polynomial trend models, we adopted a simpler approach by stratifying the data and fitting separate models for each land type.
We grouped land use into eight types: cropland, forest, grassland, shrubland, wetland, water body, impervious surface, and bare land. Province-wide land use data \citep{liu:14} is adopted for spatial predictions. Wetlands and water bodies have no soil erosion and were excluded from modeling, leaving a total of $L = 6$ land types to model.

More precisely, write $\ttv{land}_s$ as a categorical random variable whose distribution corresponds to the land type proportion of sampling unit $s$ for $l = 1, \dots, L$.
We write
$
Q_{ls} \defas  Q_s | \ttv{land}_s = l
$
so that $Q_{ls}$ represents the soil loss given land type $l$. It now suffices to model $Q_{1s}, \dots, Q_{Ls}$ separately.

At the surveyed sampling unit, the target $Q_{ls} \geq 0$ has a non-negligible mass of zero values (c.f. the lower-right panel of \autoref{fig:sample.unit}). 
While a positive point mass is not prohibited in the Wasserstein space formalism (as in \eqref{mdl:x}), it translates to phase variation of quantile functions (along the $t$-direction), which could pose difficulty for interpreting the subsequent functional principal component analysis. 
Therefore we isolate the point mass at zero using a zero-hurdled model 
\begin{equation} \label{mdl:ern_p0}
    p_{0ls} \coloneqq \prob\left( Q_{ls} = 0 \right) =
    \logit\inv (\eta_{0l}(s)),
    \quad 
    \eta_{0l}(s) = g_{0l} \circ u_l(s) + z_{0l}(s)
\end{equation}
for some trend $g_{0l}$, land type-specific covariate $u_l$ to be described shortly, and latent zero-mean stationary random field $z_{0l}$.
Conditional on $Q_{ls} > 0$, we model in the Wasserstein space the distribution of the log-erosion  $Y_{ls} \coloneqq \log Q_{ls}$ \label{mdl:ern_logscale} due to the observed right-skewness. 

We are now ready to apply the proposed model on the distribution $\Lambda^Y_{ls}$ of $Y_{ls}$.
Write
\begin{equation}\label{mdl:ern_wassPCA}
    \Lambda^Y_{ls}
    =
    \wassexp[\Lambda_{\fave, l}]\left( X_{ls} \right), \text{ with } X_{ls}(t) = \sum_{k = 1}^K  \eta_{kl}(s) \vphi_{kl}(t),
\end{equation}
where for $k = 0, 1, \dots, K$, the scalar random fields are
\begin{equation}\label{mdl:ern_wassPCA_par}
    \eta_{kl}(s) = g_{kl} \circ u_l (s) + z_{kl}(s).
\end{equation}
Here, for any land type $l$, $\Lambda_{\fave,l}$ is the spatial Fr\'echet mean as in (\ref{def:sp_frechet_mean}) over the erosion distributions for land type $l$; 
$\vphi_{1l}, \dots, \vphi_{Kl}, \dots$ are the eigenfunctions as in \eqref{eq:cov_surface_eig_problem}, 
the $u_l:\calS \to \dsR^p$ are some spatial covariates;
the $g_{kl}: \dsR^p \to \dsR$ are trends; 
the $z_{kl}$ are mean-zero stationary random fields for which the covariance structure also follows a \matern covariance model with land type-specific parameters. 
The data demonstrates almost no cross-dependency as shown in 
\suppref{Subsection S8.2} 
of the Supplement \citep{erosion:supp}, so we assume that $\cov(z_{kl}(s), z_{k'l}(s')) = 0$ for $k \not= k'$ and all $s, s' \in \calS$.
Combining \eqref{mdl:ern_rk_prod}-\eqref{mdl:ern_wassPCA_par} gives a model for local erosion distribution. 
We followed the procedure discussed in \autoref{sec:estimation} to estimate the model components. 

In sum, the proposed model is imposed on the log scale of the positive erosion, conditioning on land type and offset by \ttv{R} and \ttv{K}.
The observation at surveyed sampling unit $s$ consists of observed zero probabilities and a sample of $Y_{ls, 1}, \dots, Y_{ls, N_{ls}}$ from different land type $l$, where the local sample size $N_{ls}$ is determined by the size of surveyed location and resolution. 
Sufficiently large sample size is needed for estimating local erosion distribution, therefore sampling units stratified by land use with fewer than 50 observations (equivalently $5000$ $m^2$) were dropped from the analysis, which accounts for only less than 0.6\% of the surveyed locations.
We set $K=2$ in \eqref{mdl:ern_wassPCA}, which already accounted for more than 98\% of variation 
(see 
\suppref{Figure S10.1} 
of the Supplement \citep{erosion:supp}).
Finally, the predicted zero-probabilities and $Y$-distributions were used to derive the distributions of the extrinsic factor $Q$, which was then merged with the intrinsic factors and aggregated across land-use types to obtain erosion predictions.

\subsection{Trend and Covariates}\label{sec:covariates}

In the calculation \eqref{def:soil_loss} of the erosion value, land topography through \ttv{L} and \ttv{S} are crucial quantities. 
To improve prediction and alleviate dependence on high resolution topography, we utilized some covariates derived from openly available 30m Shuttle Radar Topography Mission (SRTM, \cite{farr:00}) digital elevation model (DEM). 
The following covariates $u$ were computed from the DEM for all sampling units and land types: The mean $\ttv{meanDEM}$ and standard deviation $\ttv{sdDEM}$ (resp.~${\ttv{meanSlope}}, {\ttv{sdSlope}}$) of the elevation (resp.~the slope), and the proportion of flat surface ${\ttv{propFlat}}$, which is the surface with slope smaller than a certain threshold. 
These covariates were then used to model the trend components in \eqref{mdl:ern_p0} and \eqref{mdl:ern_wassPCA_par}. 
Different covariates were included in different models as listed in \autoref{tbl:ern_trend_covariates}. While as many as five covariates may be included, the dense sampling and correlations among the covariates reduce the effective dimensionality, hence allowing local likelihood estimation of the trend.
See 
\suppref{Section S7} 
of the Supplement \citep{erosion:supp} for additional detail.

\begin{table}[h]
    \centering
    \caption{Covariates used for modeling trend}\label{tbl:ern_trend_covariates}
    \begin{tabular}{cc|ccccc}
        \toprule
        parameter    & mean              & \ttv{meanDEM} & \ttv{sdDEM} & \ttv{meanSlope} & \ttv{sdSlope} & \ttv{propFlat} \\
        \midrule
        $\eta_{0l}$        & $g_{0 l}$        & $\checkmark$      & $\checkmark$    & $\checkmark$        & $\checkmark$      & $\checkmark$       \\
        $\eta_{1l}$     & $g_{1l}$          & $\checkmark$      &      & $\checkmark$        &        &         \\
        $\eta_{2l}$     & $g_{2l}$          &        & $\checkmark$    & $\checkmark$        & $\checkmark$      &        
        \\ \bottomrule
    \end{tabular}
\end{table}

\subsection{Flexible Province-wide Prediction}

\begin{figure}
    \centering
    \begin{subfigure}{.55\textwidth}
        \includegraphics[width=\linewidth]{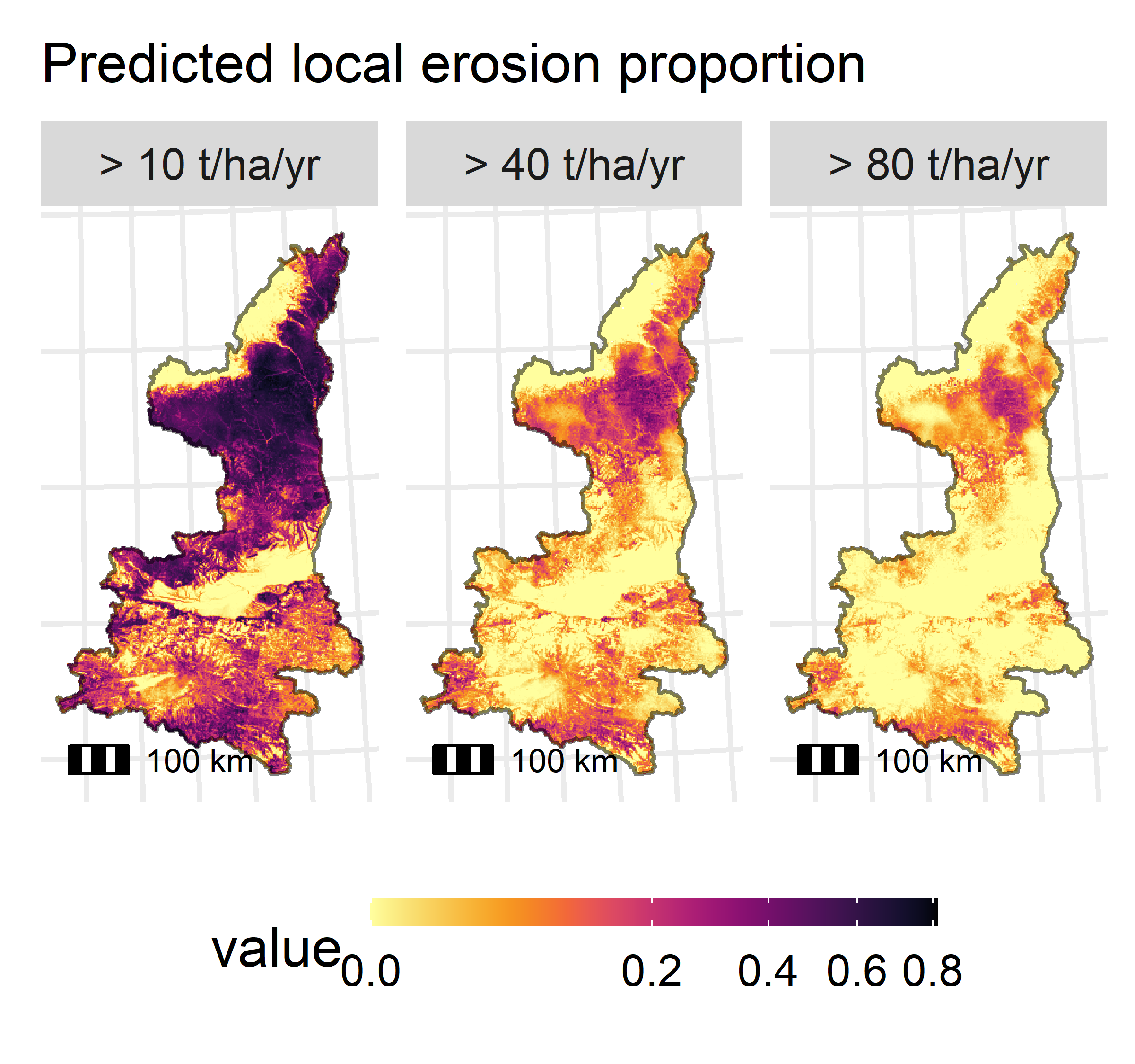}
        \caption{
            Local proportion of highly eroded lands. 
        }
        \label{fig:ern_pred_excd}
    \end{subfigure}
    \quad 
    \begin{subfigure}{.35\textwidth}
        \includegraphics[width=\linewidth]{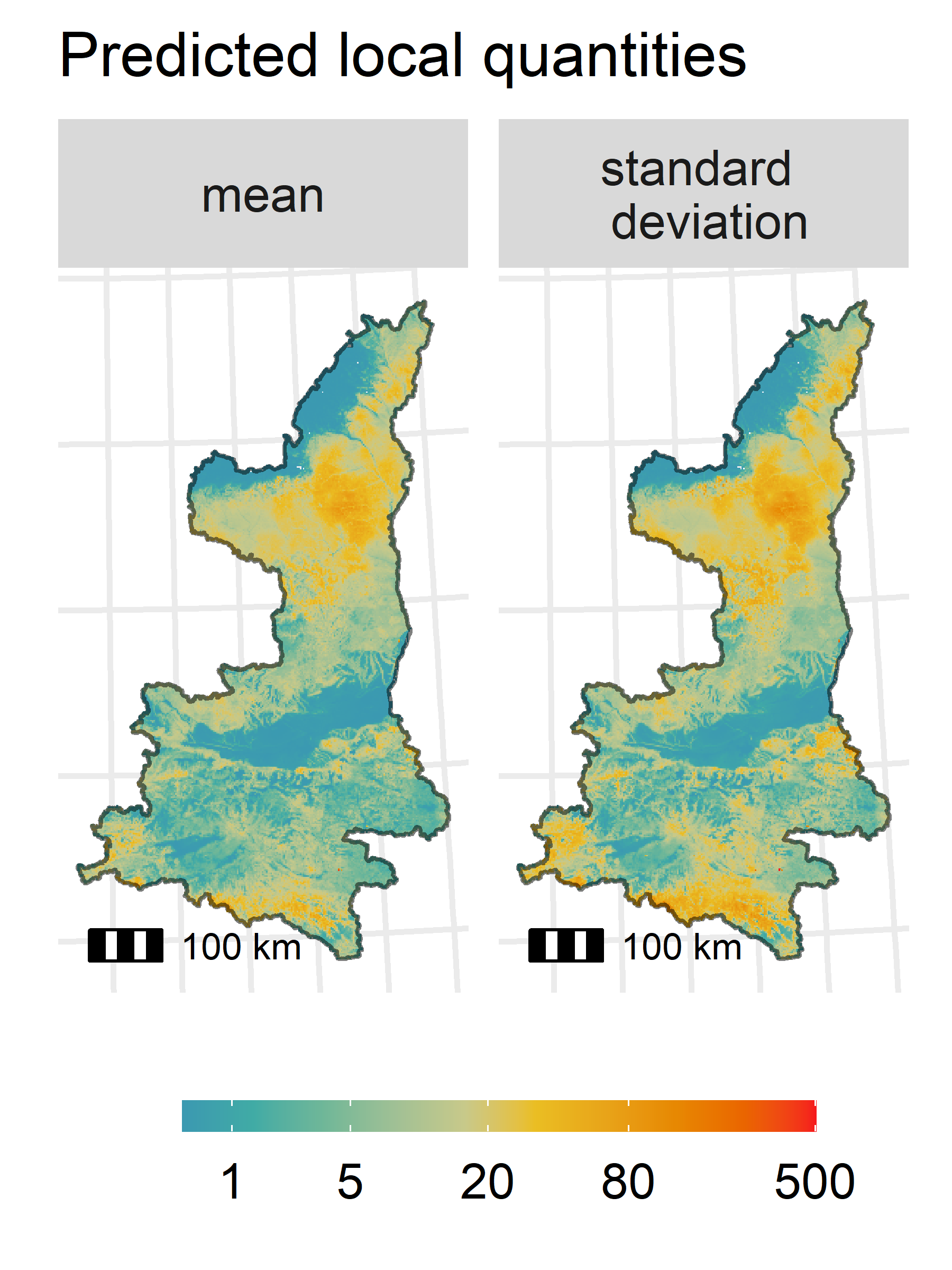}
        \caption{Local erosion mean and standard deviation. Values $\geq 500$ colored as 500.
        }
        \label{fig:ern_pred_meansd}
    \end{subfigure}
    \caption{Province-wide prediction map for local erosion distributions.}
\end{figure}

Following the previously described model, a province-wide prediction for local erosion distribution was obtained at the resolution of $1\times1$ kilometer, where for every one square kilometer pixel a predicted local erosion distribution is available. Such resolution is chosen in alignment to the surveyed sampling units.
To quantify the severity of local erosion, we inspected the exceedance probability $\prob\left(\ttv{A}_s \geq \tau\right)$ at several meaningful thresholds $\tau$.
Similar to \cite{yin:18-1}, we set the threshold $\tau$ at $5, 10, 20, 40$, and $80$ and categorized erosion into six severity levels as mild ($<$5 t/ha/yr), slight (5--10 t/ha/yr), moderate (10--20 t/ha/yr), high (20--40 t/ha/yr), severe (40--80 t/ha/yr), and extreme ($>$80 t/ha/yr). 
\autoref{fig:ern_pred_excd} displays the local proportions of highly eroded lands ($>$10, $>$40, $>$80 t/ha/yr), indicating erosion is most severe in the northern part of the province located on the loess plateau.
Moreover, the proposed method can be applied to estimate any functional of the local erosion distributions (not limited to the exceedance probability) as demonstrated in \autoref{fig:ern_pred_meansd}, which shows the predicted mean and standard deviation of the local erosion distribution. 

\begin{figure}[h]
    \centering
    \includegraphics[width=1\textwidth]{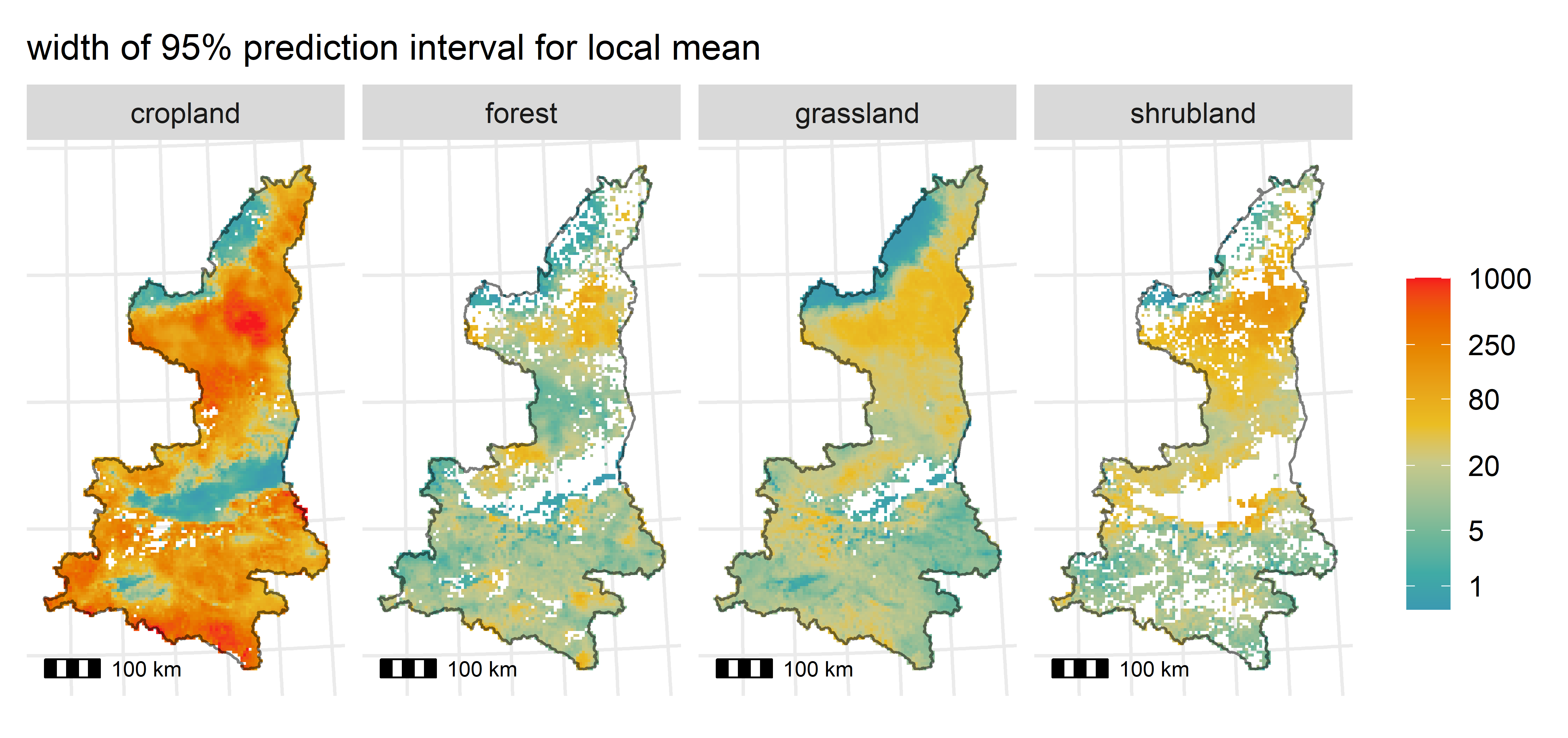}
    \caption{
        Uncertainty in predicting local erosion mean per land type, quantified by the width of 95\% prediction interval, to which the pixels are colored. Values greater than 1000 are colored as 1000. Blank pixels indicate no such land type there.
    }
    \label{fig:ern_pred_mean_uncertainty}
\end{figure}

To quantify uncertainty, \autoref{fig:ern_pred_mean_uncertainty} shows the width of 95\% prediction intervals, which were obtained following the procedure described in \autoref{sec:uncertainty} under a Gaussianity assumption (validated in \suppref{Subsection S8.3} of the Supplement \citep{erosion:supp}). We present the results conditional on the land type, showing the four most common land types.
Cropland had the greatest uncertainty in prediction, suggesting that its erosion is the hardest to predict. It has the potential for the worst erosion, presumably due to a lack of cover crop during the non-growing season.

\begin{figure}[h]
    \centering
    \includegraphics[width=\textwidth]{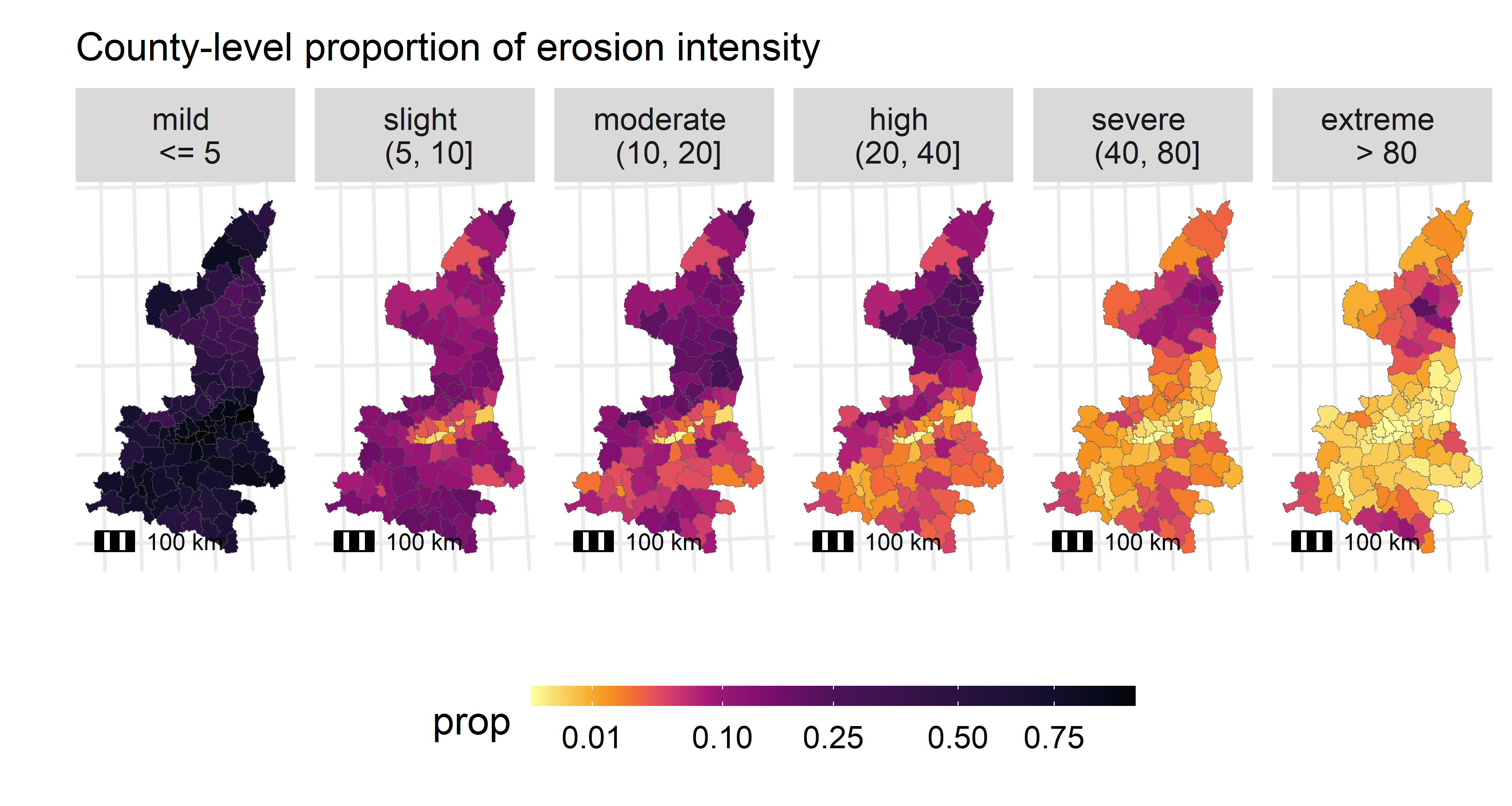}
    \caption{County-level erosion severity. Each polygon represents one county in the Shaanxi province, colored by the proportion of its land suffering a certain level of erosion.}
    \label{fig:ern_xzq}
\end{figure}

The $1\times1$ km resolution prediction can be easily aggregated into a lower resolution to tackle a larger area (e.g., county or basin) using the weighted average of the cumulative distribution functions. 
\autoref{fig:ern_xzq} demonstrates an example of erosion severity at the county level.
The simple aggregation process provides a convenient tool for administrative authorities to improve efficient treatment resource allocation across multiple watersheds.

\begin{figure}
    \centering
    \includegraphics[width=0.75\textwidth]{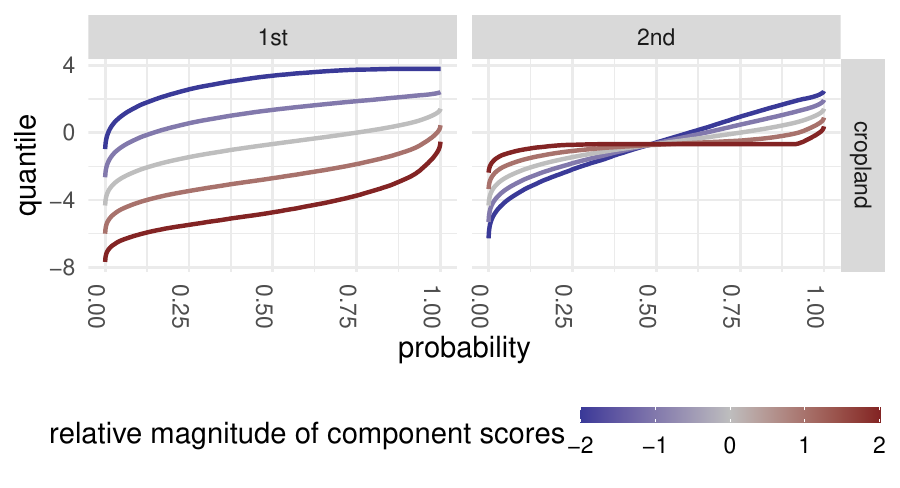}
    \caption{
        Mode of variation in terms of quantile functions of the working variable (effectively the log-erosion). The basis coefficients take values of the relative magnitude times the square root of the corresponding eigenvalue.
    }
    \label{fig:ern_mov}
\end{figure}

Close inspection of the resulting eigenfunctions $\emp \vphi_k$ suggests that the local erosion distributions vary almost within a location-scale family. \autoref{fig:ern_mov} shows the principal modes of variation plot in the quantile functions of the log-erosion.
Changes in the first component translate to a vertical shift of the quantile functions, i.e., a location variation; 
while the second component translates to flatter v.s.\ steeper quantile functions, corresponding to scale variation.
We highlight that the near location-scale family obtained from our approach is completely data-driven, as opposed to specifying a parametric family (e.g., log-normal), which we later benchmark our method against.

\subsection{Performance Assessment}\label{sec:cv}

We assessed the prediction performance in terms of accuracy and variability via repeated cross-validation.
We considered the prediction errors of local erosion distributions and the derived quantities, including the local mean, standard deviation, and exceedance probability.

The proposed model \eqref{mdl:ern_rk_prod} -- \eqref{mdl:ern_wassPCA_par} was compared to a zero-hurdled log-normal parametric model, the details of which are included in 
\suppref{Section S9} 
of the Supplement \citep{erosion:supp}.
This parametric model can be considered as a special case of the proposed approach:
apply $K = 2$ and a pre-specify basis $\vphi_{1l}(t) = 1$, $\vphi_{2l}(t) = t$ for all land use $l$ in \eqref{mdl:ern_wassPCA}, 
the $\Lambda^Y_{ls}$ is the probability measure of a normal distribution, while the coefficients $\eta_{1l}$ and $\eta_{2l}$ are functions of the mean and standard deviation.
We use the same model for the zero probabilities as in \eqref{mdl:ern_p0}, while the spatial models for the mean and standard deviation parameters of the log-normal distribution share the same forms and covariates as those for $\eta_{1l}$ and $\eta_{2l}$ in \autoref{tbl:ern_trend_covariates}, except that model \eqref{mdl:ern_wassPCA_par} is imposed on the log of the standard deviation for positive constraint.

\begin{figure}
    \centering
    \includegraphics[width=0.9\textwidth]{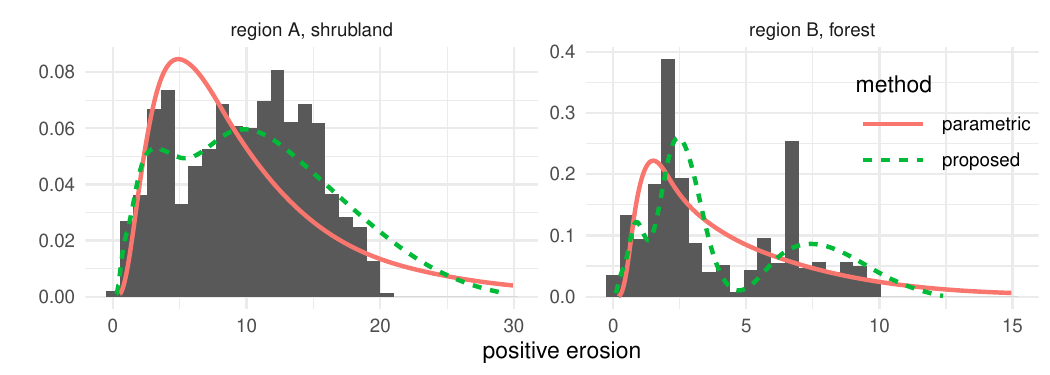}
    \caption{Comparing predicted densities of the log-normal model (solid curves) and the proposed method (dotted curves) to observed histograms in two sampling units.}
    \label{fig:ern_cv_pdf}
\end{figure}

\autoref{fig:ern_cv_pdf} compares the predicted density functions of positive erosion to the observed histograms of positive erosion in some surveyed locations in one repeat of the cross-validation process, demonstrating that the proposed method is more flexible compared to the log-normal parametric model, especially noting that the proposed method successfully captures multi-modality. 

Displayed in \autoref{tbl:ern_cv_err} are detailed comparisons of prediction errors per land type over 20 repeats of 10-fold cross-validation, which shows that the proposed method enjoys smaller errors overall, particularly in terms of local standard deviation (sd) prediction and mean Wasserstein error (MWE).
The mean absolute errors (MAE) was computed for each repeat of cross validation by averaging absolute prediction errors among all surveyed sampling units of a certain land type, the average and range (the maximum minus minimum) of which among the repeats are reported. 
Similarly, the errors of predicted local erosion distribution are quantified by average and range of MWE \eqref{def:mwe}, using the empirical quantile functions as underlying truth for computation.

To demonstrate that the distributional modeling is still valuable even when only targeting local univariate summary statistics, we also compare against a naive univariate Kriging approach in \autoref{tbl:ern_cv_err} which predicts average erosion per location based on the averages at observed locations 
(see 
\suppref{Section S9} 
of the Supplement \citep{erosion:supp} for details). Note that the performance of methods utilizing the local erosion distribution model (parametric and proposed) outperforms the naive Kriging method utilizing only the local summary statistics. Corresponding local standard deviation and MWE in \autoref{tbl:ern_cv_err} are left blank since this naive method does not produce local distribution predictions like our proposal.

\begin{table}
    \centering
    \caption{Errors in local erosion mean, standard deviation (sd), cumulative distribution function (CDF), and exceedance probability (\%) prediction over 20 repeats of 10-fold cross-validation, with range (maximum minus minimum among 20 repeats) shown in the bracket. Mean absolute errors (MAE) are shown for local mean, standard deviation, and exceedance probability, while mean Wasserstein errors (MWE) shown CDF. Each row compares the performance for a specific land use type, whereas the last row compares the performance pooling all land.}\label{tbl:ern_cv_err}
    \begin{tabular}[t]{lrrrrrr}
        \toprule
        \multicolumn{1}{c}{ } & \multicolumn{2}{c}{MAE (range) in statistics} & \multicolumn{1}{c}{error in CDF} & \multicolumn{3}{c}{MAE (range) in exceedance probability (\%)} \\
        \cmidrule(l{3pt}r{3pt}){2-3} 
        \cmidrule(l{3pt}r{3pt}){5-7}
        method & local mean & local sd & MWE (range) & >10 t/ha/yr & >40 t/ha/yr & >80 t/ha/yr\\
        \midrule
        \addlinespace[0.3em]
        \multicolumn{7}{l}{\textbf{cropland}}\\
        \cellcolor{gray!10}{\hspace{1em}Kriging} & \cellcolor{gray!10}{14.20 (0.28)} & \cellcolor{gray!10}{} & \cellcolor{gray!10}{} & \cellcolor{gray!10}{17.03 (0.23)} & \cellcolor{gray!10}{14.61 (0.21)} & \cellcolor{gray!10}{8.74 (0.14)}\\
        \hspace{1em}parametric & 14.07 (0.28) & 22.46 (0.60) & 32.65 (0.59) & 15.25 (0.28) & 12.55 (0.20) & 7.75 (0.16)\\
        \cellcolor{gray!10}{\hspace{1em}proposed} & \cellcolor{gray!10}{13.88 (0.25)} & \cellcolor{gray!10}{13.54 (0.23)} & \cellcolor{gray!10}{21.75 (0.30)} & \cellcolor{gray!10}{15.17 (0.25)} & \cellcolor{gray!10}{12.41 (0.20)} & \cellcolor{gray!10}{7.82 (0.21)}\\
        \addlinespace[0.3em]
        \multicolumn{7}{l}{\textbf{forest}}\\
        \hspace{1em}Kriging & 3.17 (0.05) &  &  & 13.37 (0.14) & 1.36 (0.00) & 0.06 (0.00)\\
        \cellcolor{gray!10}{\hspace{1em}parametric} & \cellcolor{gray!10}{3.00 (0.04)} & \cellcolor{gray!10}{2.66 (0.05)} & \cellcolor{gray!10}{4.68 (0.05)} & \cellcolor{gray!10}{11.97 (0.22)} & \cellcolor{gray!10}{1.53 (0.02)} & \cellcolor{gray!10}{0.10 (0.00)}\\
        \hspace{1em}proposed & 2.99 (0.04) & 2.45 (0.04) & 4.07 (0.05) & 11.64 (0.25) & 1.37 (0.02) & 0.06 (0.00)\\
        \addlinespace[0.3em]
        \multicolumn{7}{l}{\textbf{grassland}}\\
        \cellcolor{gray!10}{\hspace{1em}Kriging} & \cellcolor{gray!10}{3.41 (0.07)} & \cellcolor{gray!10}{} & \cellcolor{gray!10}{} & \cellcolor{gray!10}{14.99 (0.26)} & \cellcolor{gray!10}{2.20 (0.03)} & \cellcolor{gray!10}{0.02 (0.00)}\\
        \hspace{1em}parametric & 3.34 (0.13) & 3.72 (0.16) & 7.00 (0.19) & 11.81 (0.25) & 3.57 (0.13) & 0.46 (0.07)\\
        \cellcolor{gray!10}{\hspace{1em}proposed} & \cellcolor{gray!10}{3.12 (0.07)} & \cellcolor{gray!10}{2.19 (0.04)} & \cellcolor{gray!10}{4.40 (0.06)} & \cellcolor{gray!10}{10.77 (0.25)} & \cellcolor{gray!10}{2.24 (0.14)} & \cellcolor{gray!10}{0.02 (0.00)}\\
        \addlinespace[0.3em]
        \multicolumn{7}{l}{\textbf{shrubland}}\\
        \hspace{1em}Kriging & 3.78 (0.18) &  &  & 15.11 (0.40) & 3.10 (0.12) & 0.23 (0.01)\\
        \cellcolor{gray!10}{\hspace{1em}parametric} & \cellcolor{gray!10}{3.29 (0.52)} & \cellcolor{gray!10}{3.89 (2.11)} & \cellcolor{gray!10}{6.29 (0.62)} & \cellcolor{gray!10}{11.18 (0.28)} & \cellcolor{gray!10}{2.97 (0.10)} & \cellcolor{gray!10}{0.58 (0.04)}\\
        \hspace{1em}proposed & 3.09 (0.10) & 2.39 (0.06) & 4.35 (0.10) & 10.76 (0.27) & 2.54 (0.13) & 0.25 (0.02)\\
        \addlinespace[0.3em]
        \multicolumn{7}{l}{\textbf{impervious surface}}\\
        \cellcolor{gray!10}{\hspace{1em}Kriging} & \cellcolor{gray!10}{3.24 (0.04)} & \cellcolor{gray!10}{} & \cellcolor{gray!10}{} & \cellcolor{gray!10}{4.59 (0.15)} & \cellcolor{gray!10}{1.55 (0.10)} & \cellcolor{gray!10}{0.88 (0.02)}\\
        \hspace{1em}parametric & 3.20 (0.03) & 3.82 (0.09) & 5.26 (0.07) & 4.56 (0.13) & 1.50 (0.06) & 0.88 (0.01)\\
        \cellcolor{gray!10}{\hspace{1em}proposed} & \cellcolor{gray!10}{3.21 (0.04)} & \cellcolor{gray!10}{3.70 (0.09)} & \cellcolor{gray!10}{5.08 (0.10)} & \cellcolor{gray!10}{4.74 (0.23)} & \cellcolor{gray!10}{1.49 (0.06)} & \cellcolor{gray!10}{0.89 (0.02)}\\
        \addlinespace[0.3em]
        \multicolumn{7}{l}{\textbf{mix}}\\
        \hspace{1em}Kriging & 3.96 (0.05) &  &  & 10.65 (0.11) & 3.54 (0.05) & 1.41 (0.02)\\
        \cellcolor{gray!10}{\hspace{1em}parametric} & \cellcolor{gray!10}{3.74 (0.07)} & \cellcolor{gray!10}{8.87 (0.46)} & \cellcolor{gray!10}{11.50 (0.78)} & \cellcolor{gray!10}{8.82 (0.11)} & \cellcolor{gray!10}{3.09 (0.04)} & \cellcolor{gray!10}{1.33 (0.02)}\\
        \hspace{1em}proposed & 3.66 (0.05) & 5.68 (0.10) & 7.97 (0.10) & 8.54 (0.12) & 2.95 (0.06) & 1.30 (0.03)\\
        \bottomrule
    \end{tabular}
    
\end{table}

\autoref{tbl:ern_cv_err} also shows the average prediction errors of exceedance probability with respect to different thresholds compared to the naive Kriging and the parametric approach. 
This baseline method applies the classic Kriging approach on the observed exceedance probability with logistic transformation and utilizes a similar set of covariates as listed in \autoref{tbl:ern_trend_covariates} (see 
\suppref{Section S9} 
of the Supplement \citep{erosion:supp} for details).
It is clear that the proposed method outperforms these baselines consistently.

\section{Discussion}

This paper presents a novel approach to tackle local soil erosion distribution when surveyed watersheds only sparsely cover the region of interest: the Shaanxi province in China, an area profoundly impacted by erosion along the Yellow River basin. 
The proposed method addresses the limitations of existing approaches that rely on summary statistics that neglect important details in the soil loss distribution. By employing functional data analysis, our approach can predict the entire distribution at unsurveyed watersheds without imposing parametric assumptions. As a distribution-oriented method, it allows simultaneous prediction, without invoking multiple models, of various quantities of interest, such as the proportion of highly eroded lands, the mean, and the standard deviation of local erosion. 
Furthermore, our approach accommodates nonlinear covariates and accounts for spatially varying predictors while also providing uncertainty quantification.
The computation is made affordable by employing dimension reduction techniques, enabling predictions of soil erosion distribution at high spatial resolution, which is crucial for informed policy making. 
Our methodology offers a versatile approach to spatial prediction of distribution data and their functionals, enabling more accurate and efficient analysis for various spatially distributed distributional data types. 


\begin{acks}[Acknowledgments]
    The authors would like to thank the anonymous referees, the Associate
    Editor and the Editor for their careful review and constructive comments that improved the quality of this paper.
\end{acks}

\begin{funding}

Dr.~Dai was partially supported by NSF grant DMS-2113713.
Dr.~Zhu's research is partially supported by U.S.~Department of Agriculture's National Resources Inventory, Cooperative Agreement NR203A750023C006, Great Rivers CESU 68-3A75-18-504, and NSF CBET 2041859.

\end{funding}

\begin{supplement}
\stitle{Supplement A}
\sdescription{Includes additional practical considerations, simulation studies, and some theoretical derivation.}
\end{supplement}
\begin{supplement}
\stitle{Supplement B}
\sdescription{Provides an R package skeleton (code and documentation) implementing the proposed method. It is also available through the GitHub repository: \href{https://github.com/jiamingqiu/spdf}{https://github.com/jiamingqiu/spdf}.}
\end{supplement}


\bibliographystyle{imsart-nameyear} 
\bibliography{usedrefs}

\end{document}